\documentclass[sigconf]{acmart}
\usepackage{multirow}
\usepackage{float}
\usepackage{subfig}
\usepackage{enumitem}
\usepackage{algorithmic}
\usepackage[ruled, linesnumbered]{algorithm2e}

\AtBeginDocument{%
  \providecommand\BibTeX{{%
    \normalfont B\kern-0.5em{\scshape i\kern-0.25em b}\kern-0.8em\TeX}}}

\setcopyright{acmlicensed}

\copyrightyear{2024}
\acmYear{2024}
\setcopyright{acmlicensed}\acmConference[KDD '24]{Proceedings of the 30th ACM SIGKDD Conference on Knowledge Discovery and Data Mining}{August 25--29, 2024}{Barcelona, Spain}
\acmBooktitle{Proceedings of the 30th ACM SIGKDD Conference on Knowledge Discovery and Data Mining (KDD '24), August 25--29, 2024, Barcelona, Spain}
\acmDOI{10.1145/3637528.3671824}
\acmISBN{979-8-4007-0490-1/24/08}

\begin{document}

\newcommand{\shortname}{\emph{PAAC}}
\newcommand{\fullname}{\emph{\underline{P}opularity-\underline{A}ware \underline{A}lignment and \underline{C}ontrast ~ (PAAC)}}
\newcommand{\bhy}[1]{{\textcolor{blue}{[ #1]}}}

\title{Popularity-Aware Alignment and Contrast for Mitigating Popularity Bias}

\author{Miaomiao Cai}
\affiliation{%
  \institution{Hefei University of Technology}
  \city{Hefei}
  \country{China}
}
\email{cmm.hfut@gmail.com}

\author{Lei Chen}
\affiliation{%
  \institution{Tsinghua University}
  \city{Beijing}
  \country{China}
}
\email{chenlei.hfut@gmail.com}

\author{Yifan Wang}
\affiliation{%
  \institution{DCST, Tsinghua University}
  \country{China}
  \city{Beijing}
}
\email{yf-wang21@mails.tsinghua.edu.cn}

\author{Haoyue Bai}
\affiliation{%
  \institution{Hefei University of Technology}
  \city{Hefei}
  \country{China}
}
\email{baihaoyue621@gmail.com}

\author{Peijie Sun}
\affiliation{%
  \institution{DCST, Tsinghua University}
  \city{Beijing}
  \country{China}
}
\email{sun.hfut@gmail.com}

\author{Le Wu}
\affiliation{%
  \institution{Hefei University of Technology}
  \city{Hefei}
  \country{China}
}
\email{lewu.ustc@gmail.com}

\author{Min Zhang}
\authornote{Corresponding Authors.}
\affiliation{%
    \institution{DCST, Tsinghua University}
    \institution{Quan Cheng Laboratory}
    \city{Beijing}
    \country{China}
}
\email{z-m@tsinghua.edu.cn}

\author{Meng Wang}
\authornotemark[1]
\affiliation{%
  \institution{Hefei University of Technology}
  \city{Hefei}
  \country{China}
}
\email{eric.mengwang@gmail.com}

\renewcommand{\shortauthors}{Miaomiao Cai, et al.}

\begin{abstract}
Collaborative Filtering~(CF) typically suffers from the significant challenge of popularity bias due to the uneven distribution of items in real-world datasets. 
This bias leads to a significant accuracy gap between popular and unpopular items.
It not only hinders accurate user preference understanding but also exacerbates the Matthew effect in recommendation systems. 
To alleviate popularity bias, existing efforts focus on emphasizing unpopular items or separating the correlation between item representations and their popularity.
Despite the effectiveness, existing works still face two persistent challenges: (1) how to extract common supervision signals from popular items to improve the unpopular item representations, and (2) how to alleviate the representation separation caused by popularity bias.
In this work, we conduct an empirical analysis of popularity bias and propose ~\fullname ~ to address two challenges. 
Specifically, we use the common supervisory signals modeled in popular item representations and propose a novel popularity-aware supervised alignment module to learn unpopular item representations. 
Additionally, we suggest re-weighting the contrastive learning loss to mitigate the representation separation from a popularity-centric perspective.
Finally, we validate the effectiveness and rationale of ~\shortname ~ in mitigating popularity bias through extensive experiments on three real-world datasets.
Our code is available at https://github.com/miaomiao-cai2/KDD2024-PAAC.

\end{abstract}



\maketitle

\section{Introduction}

Modern recommender systems play a crucial role in mitigating information overload~\cite {chen2019efficient, sun2024collaborative, covington2016deep,Brain_signal,Bottleneced}. 
Collaborative filtering ~(CF) is widely used in personalized recommendations to help users find items of potential interest.
CF-based methods primarily learn user preferences and item characteristics by aligning the representations of users and the items they interact with~\cite{citationsurveylekey, Koren2009MatrixFT}.
Despite their success, CF-based methods often face popularity bias~\cite{Chen2020BiasAD,wang2023survey}, resulting in significant accuracy gaps between popular and unpopular items~\cite{Wei2020ModelAgnosticCR,Zhu2021PopularityOpportunityBI}.
Popularity bias stems from the limited supervisory signals for unpopular items, causing overfitting during training and reducing performance on the test set.
It hinders the accurate understanding of user preferences, decreasing recommendation diversity~\cite{AutoDebias,shao2022faircf,jiang2024item}.
What's even worse is that popularity bias may exacerbate the Matthew effect, where popular items become even more popular due to frequent recommendations~\cite{wang2023survey, Zhu2021PopularityBI}.

\begin{figure}
    \centering
    \subfloat{\includegraphics[width =0.555\linewidth]{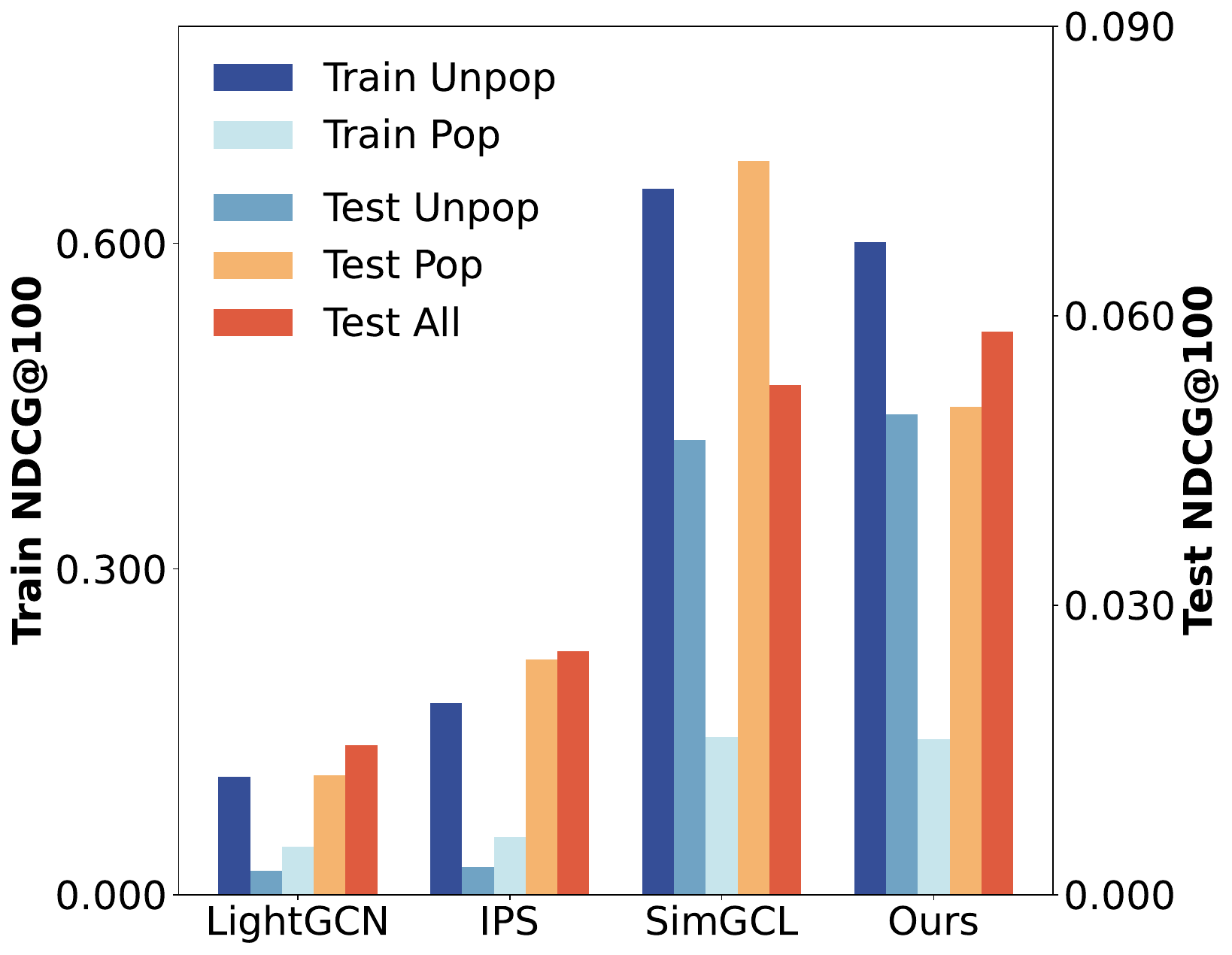}\label{fig:intro_overfitting}}
    \hfill
    \subfloat{\includegraphics[width =0.445\linewidth]{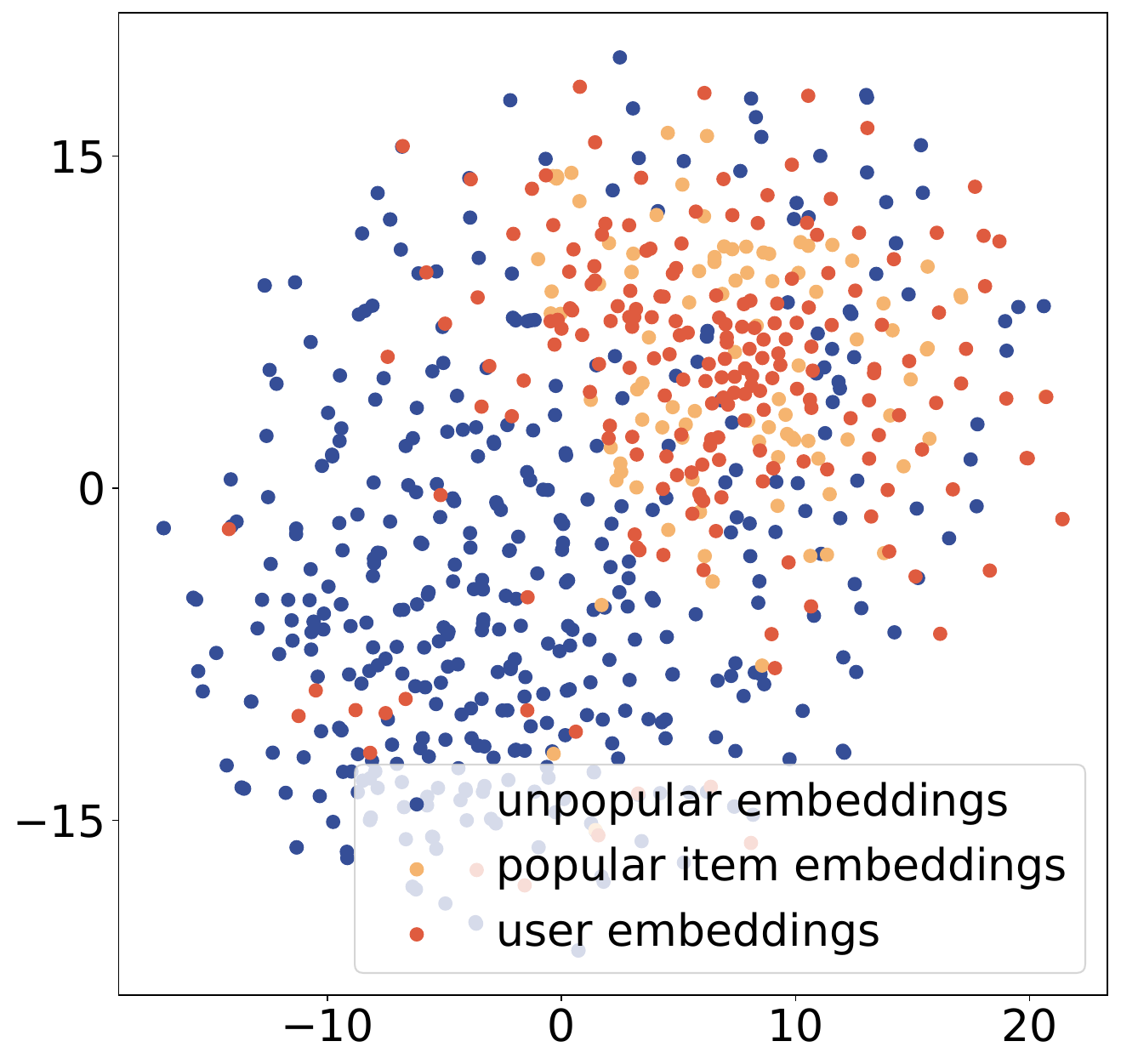}\label{fig:intro_representation}}
    \hfill
    \caption{Popularity bias presents two challenges: (1) Overfitting caused by limited supervisory signals for unpopular items, and (2) Representation separation in item embeddings driven by popularity bias.}
    \label{fig:intro}
    \vspace{-0.4cm}
\end{figure}

Mitigating popularity bias in recommendation systems, as described in Figure.~\ref{fig:intro}, presents two primary and significant challenges.
The first challenge arises from insufficient representations of unpopular items during training, leading to overfitting and poor generalization performance. 
The second challenge, representation separation, occurs when popular and unpopular items are modeled into different semantic spaces, exacerbating bias and reducing recommendation accuracy.
Next, we will explore these challenges and discuss potential solutions to mitigate popularity bias.

Due to the limited supervisory signals for unpopular items, their representations are insufficient leading to overfitting.
During training, representation alignment focuses on users and the items they have interacted with~\cite{Wang2022TowardsRA,Rendle2009BPRBP}.
However, due to limited interactions, unpopular items are often modeled around a small number of users.
This focused modeling can lead to overfitting due to the insufficient representations of unpopular items.
As shown in the left part of Figure.\ref{fig:intro}, we divide items into popular and unpopular groups based on the Pareto principle~\cite{yang2022hicf}. And then evaluate their performance in both training and testing sets (measured using $NDCG@100$).
The results reveal that traditional methods achieve higher accuracy for unpopular items during training but significantly lower accuracy during testing, indicating clear overfitting.
To address this issue, previous studies have tried to boost the training weights or prediction scores for unpopular items, such as IPS~\cite{Zhu2021PopularityOpportunityBI}, MACR~\cite{Wei2020ModelAgnosticCR}, and others~\cite{Zhao2022InvestigatingAP,AutoDebias,chen2023improving}. However, as shown in Figure.\ref{fig:intro}, overfitting still exists even with augmented supervisory weights for unpopular items.
This may be because unpopular items still lack sufficient supervisory signals, leading to inadequate representation capability.
Therefore, \textbf{how to enhance the representation modeling of unpopular items} remains a challenge.

Recent studies indicate that popularity bias causes representation separation in item embeddings~\cite{zhang2023mitigating, Yu2021AreGA}. 
Specifically, the model represents popular and unpopular items in different semantic spaces according to their popularity levels.
As shown in the right part of Figure.\ref{fig:intro}, we train LightGCN\cite{He2020LightGCNSA} on the Yelp2018 dataset\footnote{https://www.yelp.com/dataset}, randomly selected users and items, and visualize them using t-SNE~\cite{van2008visualizing} dimensionality reduction.
The blue dots represent unpopular items, the yellow dots represent popular items, and the orange dots represent users.
As seen, there is a clear distinction in the positions of unpopular and popular items in the representation space.
User representations show a preference for popular items, exacerbating popularity bias.
Existing methods try to alleviate representation separation by either removing the correlation between item representations and popularity~\cite{Wei2020ModelAgnosticCR,wu2021learning,bonner2018causal} or enhancing overall consistency through contrastive learning~\cite{wang2020understanding,Yu2021AreGA}.
However, blindly removing popularity information can harm recommendation accuracy~\cite{AutoDebias}.
While contrastive learning methods improve recommendation performance, they often worsen representation separation by pushing positive and negative samples apart.
When negative samples follow the popularity distribution~\cite{chen2020efficient,Yu2021AreGA}, most are popular items. Optimizing for unpopular items as positive samples pushes popular items further away, intensifying representation separation. Conversely, when negative samples follow a uniform distribution~\cite{Yao2020SelfsupervisedLF}, most are unpopular items. Optimizing for popular items as positive samples separates them from most unpopular items, again worsening representation separation. 
Therefore, \textbf{how to effectively solve representation separation} is also crucial.

In this work, we conduct a analysis of popularity bias and propose ~\fullname ~ to address two challenges. 
Our model ~\shortname ~  primarily consists of the following two modules:
(1) \textbf{Supervised Alignment Module}: To enhance the representations of unpopular items with more supervision signals, we use common supervisory signals modeled in popular item representations and propose a popularity-aware supervised alignment module. 
Intuitively, items interacted with by the same user share similar characteristics. 
By leveraging similar characteristics modeled in popular item representation, we propose to align the representations of popular and unpopular items interacted with by the same user.
(2) \textbf{Re-weighting Contrast Module:} To better alleviate representation separation, we propose a re-weighting contrast module from a popularity-centric perspective.
Considering the influence of various popularity levels on recommendation performance as positive and negative samples, we introduce hyperparameters $\gamma$ and $\beta$ to control the weighting of samples with different item popularity levels.
Our contributions can be summarized in three key points:
\begin{itemize}[leftmargin=0.5cm, itemindent=0cm]
    \item To provide more supervisory signals for unpopular items, we leverage common characteristics modeled in popular item representations and propose a popularity-aware supervised alignment module to enhance the unpopular item representations.
    \item To more effectively alleviate representation separation, we propose a re-weighting contrast module from a popularity-centric perspective, re-weighting the positive and negative samples.
    \item Extensive experiments on three real-world datasets demonstrate the effectiveness and rationale of ~\shortname ~ in mitigating popularity bias.
\end{itemize}

\section{preliminary}

\subsection{Collaborative Filtering}
The core of CF-based models is to learn user preferences and item characteristics by aligning user and item representations based on their interactions~\cite{Wang2022TowardsRA,chen2021set2setrank}.
Based on these representations, the trained model predicts potential interactions for recommendation~\cite{Lin2022ImprovingGC}.
Specifically, let $\boldsymbol{U}$ ($|\boldsymbol{U}|=M$) and $\boldsymbol{I}$ ($|\boldsymbol{I}|=N$) represent the sets of users and items, respectively. 
In the implicit feedback setting, the observed interactions are represented by the matrix $\mathcal{R}\in{0,1}^{M \times N}$, where $\mathcal{R} _{u,i}=1$ indicates an interaction between user $u$ and item $i$, and $\mathcal{R}_{u,i}=0$ indicates no interaction.
To better learn user preferences and item characteristics, we use LightGCN~\cite{He2020LightGCNSA} as the encoder. 
It employs Graph Convolution Networks~(GCNs) to learn high-order collaborative signals~\cite{Chen2020RevisitingGB,Wang2019NeuralGC}, mapping user/item IDs to the user representation matrix $\mathbf{Z}\in \mathbb{R}^{M \times D}$ and the item representation matrix $\mathbf{H}\in \mathbb{R}^{N \times D}$.
Next, the prediction score estimates how likely user $u$ will prefer item $i$ based on these representations.
We use the dot product~\cite{Rendle2009BPRBP,chen2021set2setrank} to define the prediction score: $s(u,i)= \mathbf{z}_u^T \mathbf{h}_i$, where $s(u,i)$ denotes the prediction score for user $u$ on item $i$, and $\mathbf{z}_u$ and $\mathbf{h}_i$ denote the representations of user $u$ and item $i$, respectively.

To better optimize the learning of representations, many studies use the Bayesian Personalized Ranking (BPR) loss~\cite{Rendle2009BPRBP}, a well-designed pairwise ranking objective for the recommendation. We apply this as the main loss for the recommendation task:
\begin{equation}
    \mathcal{L}_{rec}= -\frac{1}{|\mathcal{R}|}\sum_{(u,i,j)\in \mathcal{O^+}} ln\sigma(s(u,i)-s(u,j)),
    \label{bprloss}
\end{equation}
where $\sigma(\cdot)$ is the sigmoid function, $\mathcal{O^+}=\{(u, i,j)|\mathcal{R}_{u, i}=1,\mathcal{R}_{u,j}=0\}$ represents pairwise data, and $j$ is a randomly sampled negative item that the user has not interacted with.
 
\subsection{Contrastive Learning based CF}

Recent studies on Contrastive Learning~(CL)-based recommender systems suggest that optimizing the uniformity of items can mitigate popularity bias to some extent~\cite{jaiswal2020survey,khosla2020supervised,Lin2022ImprovingGC,Yao2020SelfsupervisedLF}. 
Specifically, CL-based models use Information Normalized Cross Entropy (InfoNCE~\cite{oord2018representation}) to minimize the distance between positive samples and maximize the distance from negative samples:
\begin{equation}
        \mathcal{L}_{cl}=\sum_{i \in I}\log\frac{exp(h_i'h_i''/\tau)}{\sum_{j\in I}exp(h_i'h_j''/\tau)},
\end{equation}
where $i$ and $j$ represent positive and negative items respectively, $h'$ and $h''$ are the item representations after different data augmentations, and $\tau>0$ is the temperature coefficient. 
In this work, we use noise perturbation for data augmentation, a simpler and more effective method than graph augmentation~\cite{Yu2021AreGA,yang2023generative}.
Although effective, CL-based methods tend to exacerbate representation separation by increasing the distance between positive and negative samples, as illustrated in Section 1.

\section{The proposed model}

\begin{figure*}
    \vspace{-0.4cm}
    \centering
    \includegraphics[width=1.0\linewidth]{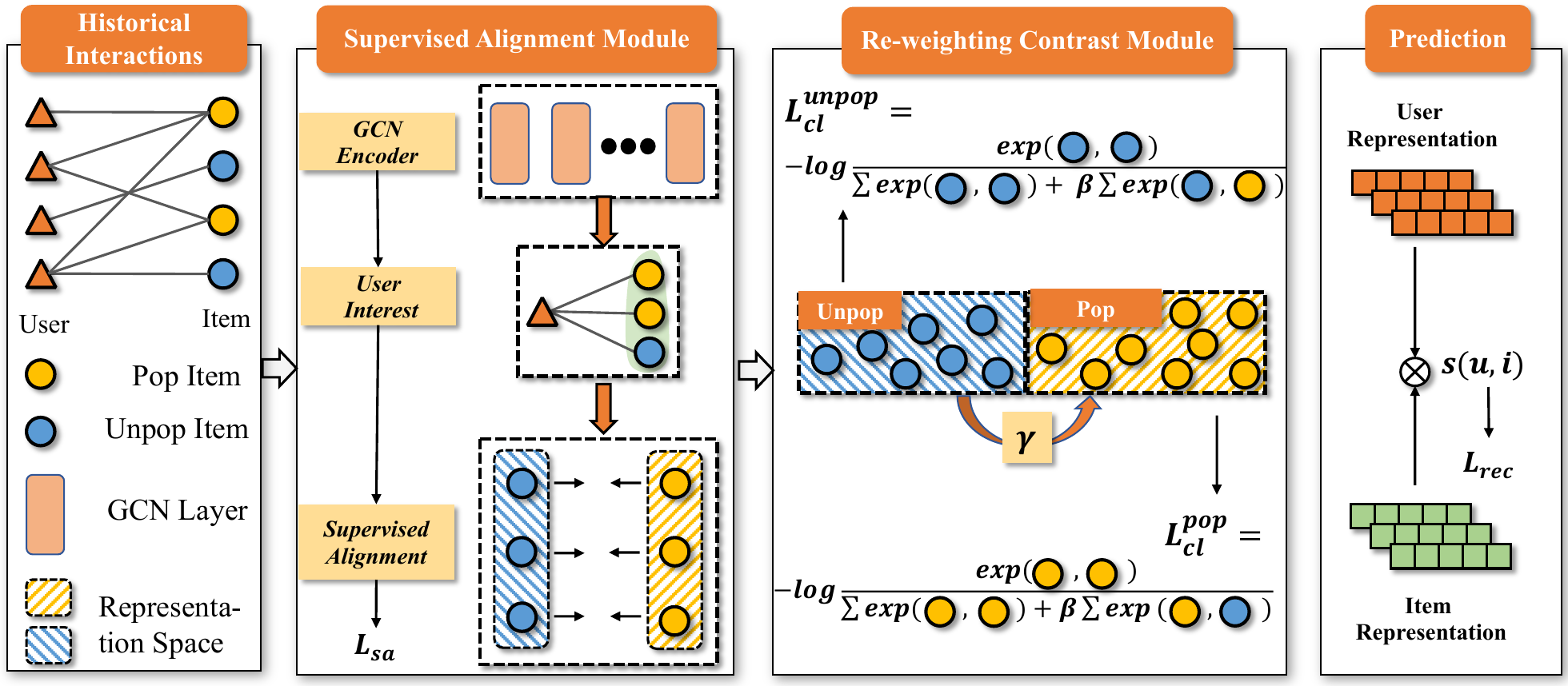}
    \caption{An Illustration of our proposed ~\fullname, which consists of the Supervised Alignment Module and the Re-weighting Contrast Module. Supervised Alignment Module leverages the common supervision signal in popular representations to guide the learning of unpopular representations. Re-weighting Contrast Module address representation separation from a popularity-centric perspective.}
    \label{fig:framework}
\end{figure*} 

To address the existing challenges in mitigating popularity bias, we propose ~\fullname, as illustrated in Figure.~\ref{fig:framework}. 
We leverage the common supervisory signals in popular item representations to guide the learning of unpopular representations and propose a popularity-aware supervised alignment module. Additionally, we introduce a re-weighting mechanism in the contrastive learning module to address representation separation from a popularity-centric perspective.

\subsection{Supervised Alignment Module}

During training, the alignment of representations typically emphasizes users and items that have interacted~\cite{Wang2022TowardsRA, Rendle2009BPRBP}, often resulting in items being closer to interacted users than non-interacted ones in the representation space.
However, due to the limited interactions of unpopular items, they tend to be modeled based on a small subset of users.
This narrow focus might lead to overfitting, as the representations of unpopular items may not adequately capture their characteristics.
As illustrated in Section 1, how to enhance unpopular representation modeling remains a challenge.

The difference in the number of supervisory signals is crucial in learning representations for popular and unpopular items. 
In particular, popular items benefit from an abundance of supervisory signals throughout the alignment process, facilitating effective learning of their representations. 
In contrast, unpopular items with a limited number of supervised users are more prone to overfitting.
This is due to insufficient representation learning for unpopular items, highlighting the impact of supervisory signal distribution on representation quality.
Intuitively, items interacted with by the same user share some similar characteristics.
In this part, we leverage common supervisory signals in popular item representations and introduce a popularity-aware supervised alignment method to enhance unpopular item representations.

Specifically, we first filter items with similar characteristics based on the user’s interests.
For any user $u$, we refer to the set of items they interact with as $I_u$:
\begin{equation}
    \boldsymbol{I}_u=\{i|i\in \boldsymbol{I} ~and~ \mathcal{R}_{u,i}=1|u\}.
\end{equation}

Consistent with prior work~\cite{zhang2023invariant,Zhu2021PopularityOpportunityBI}, we count the frequency $p(i)$ of each item $i$ appearing in the training dataset as its popularity.
Afterward, we group $I_u$ based on the relative popularity of the items $p(i)$.
For a clearer explanation, we divide $\boldsymbol{I}_u$ into two groups: the popular item group $\boldsymbol{I}_u^{pop}$ and the unpopular item group $\boldsymbol{I}_u^{unpop}$:
\begin{equation}
    \begin{aligned}
    & \boldsymbol{I}_u=\boldsymbol{I}_u^{pop} \cup \boldsymbol{I}_u^{unpop}, \\
    & \forall ~{i}\in I_u^{pop}~and~{i'}\in I_u^{unpop}, ~p(i)>p(i'),
    \end{aligned}
    \label{eqa:iu}
\end{equation}
where $\boldsymbol{I}_u^{pop}$ and $\boldsymbol{I}_u^{unpop}$ are disjoint, and the popularity of each item in the popular group is greater than that of any item in the unpopular group~\cite{liu2023mitigating}.
This means that popular items receive more supervisory information than unpopular items, leading to poorer recommendation performance for unpopular items.

To address the challenge of inadequate representation learning for unpopular items, we leverage the assumption that items interacted with by the same user exhibit some similar characteristics.
Specifically, we use similar supervisory signals in popular item representations to enhance the representations of unpopular items.
Inspired by previous works ~\cite{wang2020understanding, zhang2022correct}, we align the representations of items in $\boldsymbol{I}_u^{pop}$ and $\boldsymbol{I}_u^{unpop}$ to provide more supervisory information to unpopular items and enhance its representation, as follows:
\begin{equation}
    \mathcal{L}_{sa}=\sum_{u \in \mathcal{U}}\frac{1}{|\boldsymbol{I}_u|}\sum_{i\in \boldsymbol{I}^{pop}_u,i' \in \boldsymbol{I}^{unpop}_u}\|f(i)-f(i')\|^2,
    \label{sa}
\end{equation}
where $f(\cdot)$ is a recommendation encoder and $\mathbf{h}_i=f(i)$.
By efficiently using the inherent information in the data, we provide more supervisory signals for unpopular items without introducing additional side information.
This module enhances the representation of unpopular items, mitigating the overfitting issue.

\subsection{Re-weighting Contrast Module}

Recent studies have highlighted that popularity bias often results in a distinct representation separation of item embeddings~\cite{zhang2023mitigating, Yu2021AreGA}. 
While CL-based methods aim to enhance overall uniformity by pushing negative samples, their current sampling strategies may inadvertently worsen this separation.
When negative samples follow the popularity distribution~\cite{chen2020efficient,Yu2021AreGA}, dominated by popular items, optimizing for unpopular items as positive samples enlarges the gap between popular and unpopular items in the representation space.
Conversely, when negative samples follow a uniform distribution~\cite{Yao2020SelfsupervisedLF}, focusing on popular items separates them from the majority of unpopular ones, worsening the representation gap.
Existing studies ~\cite{Wu2020SelfsupervisedGL, Yu2021AreGA} use the same weights for positive and negative samples in the contrastive loss function, without considering differences in item popularity.
However, in real-world recommendation datasets, the impact of items varies due to dataset characteristics and interaction distributions.
Neglecting this aspect could lead to suboptimal results and exacerbate representation separation.

Inspired by previous works~\cite{park2023toward, Zhu2021PopularityOpportunityBI}, we propose to identify different influences by re-weighting different popularity items.
To this end, we introduce re-weighting different positive and negative samples to mitigate representation separation from a popularity-centric perspective.
We incorporate this approach into contrastive learning to better optimize the consistency of representations.
Specifically, we aim to reduce the risk of pushing items with varying popularity further apart.
For example, when using a popular item as a positive sample, our goal is to avoid pushing unpopular items too far away.
Thus, we introduce two hyperparameters to control the weights when items are considered positive and negative samples.

To ensure balanced and equitable representations of items within our model, we first propose a dynamic strategy to categorize items into popular and unpopular groups for each mini-batch.
Instead of relying on a fixed global threshold, which often leads to the overrepresentation of popular items across various batches, we implement a hyperparameter $x$. 
This hyperparameter readjusts the classification of items within the current batch.
By adjusting the hyperparameter $x$, we maintain a balance between different item popularity levels. This enhances the model’s ability to generalize across diverse item sets by accurately reflecting the popularity distribution in the current training context.
Specifically, we denote the set of items within each batch as $\boldsymbol{I}_B$. And then we divide $\boldsymbol{I}_B$ into a popular group $\boldsymbol{I}^{pop}$ and an unpopular group $\boldsymbol{I}^{unpop}$ based on their respective popularity levels, classifying the top $x\%$ of items as $\boldsymbol{I}^{pop}$:
\begin{equation}
    \label{eqa:ib}
    \begin{aligned}
    & \boldsymbol{I}_B=\boldsymbol{I}^{pop} \cup \boldsymbol{I}^{unpop}, \\
    & \forall ~{i}\in \boldsymbol{I}^{pop}~and~{i'}\in \boldsymbol{I}^{unpop}, ~ p(i)~>~p(i'),
    \end{aligned}
\end{equation}
where $\boldsymbol{I}^{pop} \in \boldsymbol{I}_B$ and $\boldsymbol{I}^{unpop} \in \boldsymbol{I}_B$ are disjoint, with $\boldsymbol{I}^{pop}$ consisting of the top $ x\% $ of items in the batch. 
In this work, we dynamically divided items into popular and unpopular groups within each mini-batch based on their popularity, assigning the top 50\% as popular items and the bottom 50\% as unpopular items. This radio not only ensures equal representation of both groups in our contrastive learning but also allows items to be classified adaptively based on the batch's current composition.

After that, we use InfoNCE~\cite{oord2018representation} to optimize the uniformity of item representations~\cite{wang2020understanding}. 
Unlike traditional CL-based methods, we calculate the loss for different item groups. 
Specifically, we introduce the hyperparameter $\gamma$ to control the positive sample weights between popular and unpopular items, adapting to varying item distributions in different datasets:
\begin{equation}    
\mathcal{L}_{cl}^{item}=\gamma*\mathcal{L}_{cl}^{pop}+(1-\gamma)*\mathcal{L}_{cl}^{unpop},
\label{equa:cl_group}
\end{equation}
where $\mathcal{L}_{cl}^{pop}$ represents the contrastive loss when popular items are considered as positive samples, and $\mathcal{L}_{cl}^{unpop}$ represents the contrastive loss when unpopular items are considered as positive samples.
The value of $\gamma$ ranges from 0 to 1, where $\gamma=0$ means exclusive emphasis on the loss of popular items $\mathcal{L}_{cl}^{pop}$, and $\gamma=1$ means exclusive emphasis on the loss of unpopular items $\mathcal{L}_{cl}^{unpop}$.
By adjusting $\gamma$, we can effectively balance the impact of positive samples from both popular and unpopular items, allowing adaptability to varying item distributions in different datasets.

Following this, we fine-tune the weighting of negative samples in the contrastive learning framework using the hyperparameter $\beta$. 
This parameter controls how samples from different popularity groups contribute as negative samples. 
Specifically, we prioritize re-weighting items with popularity opposite to the positive samples, mitigating the risk of excessively pushing negative samples away and reducing representation separation. Simultaneously, this approach ensures the optimization of intra-group consistency.
For instance, when dealing with popular items as positive samples, we separately calculate the impact of popular and unpopular items as negative samples.
The hyperparameter $\beta$ is then used to control the degree to which unpopular items are pushed away. 
This is formalized as follows:
{\small \begin{equation}
    \mathcal{L}_{cl}^{pop}=\sum\limits_{i\in I^{pop}}\log\frac{exp(h_i'h_i''/\tau)}{\sum\limits_{j\in I^{pop}}exp(h_i'h_j''/\tau)+\beta\sum\limits_{j\in I^{unpop}}exp(h_i'h_j''/\tau)},
\end{equation}}
similarly, the contrastive loss for unpopular items is defined as:
{\small \begin{equation}
    \mathcal{L}_{cl}^{unpop}=\sum\limits_{i\in I^{unpop}}\log\frac{exp(h_i'h_i''/\tau)}{\sum\limits_{j\in I^{unpop}}exp(h_i'h_j''/\tau)+\beta\sum\limits_{j\in I^{pop}}exp(h_i'h_j''/\tau)},
\end{equation}}
where the parameter $\beta$ ranges from 0 to 1, controlling the negative sample weighting in the contrastive loss.
When $\beta=0$, it means that only intra-group uniformity optimization is performed. 
Conversely, when $\beta=1$, it means equal treatment of both popular and unpopular items in terms of their impact on positive samples. 
The setting of $\beta$ allows for a flexible adjustment between prioritizing intra-group uniformity and considering the impact of different popularity levels in the training.
We prefer to push away items within the same group to optimize uniformity. This setup helps prevent over-optimizing the uniformity of different groups, thereby mitigating representation separation.

The final re-weighting contrastive objective is the weighted sum of the user objective and the item objective:
\begin{equation}
    \mathcal{L}_{cl}=\frac{1}{2}\times(\mathcal{L}_{cl}^{item}+\mathcal{L}_{cl}^{user}).
    \label{cl_loss_all}
\end{equation}
In this way, we not only achieved consistency in representation but also reduced the risk of further separating items with similar characteristics into different representation spaces, thereby alleviating the issue of representation separation caused by popularity bias.

\subsection{Model Optimization}
To alleviate popularity bias in collaborative filtering tasks, we utilize a multi-task training strategy~\cite{Wu2020SelfsupervisedGL} to jointly optimize the classic recommendation loss ($cf.$ Equation~(\ref{bprloss})), supervised alignment loss ($cf.$ Equation~(\ref{sa})), and re-weighting contrast loss ($cf.$ Equation~(\ref{cl_loss_all})).
\begin{equation}
\mathcal{L}=\mathcal{L}_{rec}+\lambda_{1}\mathcal{L}_{sa}+\lambda_{2}\mathcal{L}_{cl}+\lambda_3\|\Theta\|_2^2,
\label{overall_equa}
\end{equation}
where  $\Theta$ is the set of model parameters in $\mathcal{L}_{rec}$ as we do not introduce additional parameters, $\lambda{1}$ and $\lambda_{2}$ are hyperparameters that control the strengths of the popularity-aware supervised alignment loss and the re-weighting contrastive learning loss respectively, and $\lambda_3$ is the $L_2$ regularization coefficient.
After completing the model training process, we use the dot product to predict unknown preferences for recommendations. Algorithm~\ref{alg: paac} shows the detailed algorithm of ~\shortname.

\begin{algorithm} [t]
\renewcommand{\algorithmicrequire}{\textbf{Input:}}
\renewcommand\algorithmicensure {\textbf{Output:}}
\caption{\small{The Algorithm of \shortname}}\label{alg: paac}
\begin{algorithmic}[1]
\REQUIRE user-item interactions $\mathcal{R}$, recommendation encoder $f(\cdot)$, learning rate $\eta$, hyperparameters $\lambda_1$, $\lambda_2$, $\lambda_3$, $\beta$, $\gamma$;  ~~\\
\ENSURE recommendation encoder $f(\Theta)$; ~~\\
\STATE randomly initialize recommendation encoder parameter $\Theta$; \\
\FOR{$epoch=1,2,...,T$} 
    \FOR{batch data $\mathcal{B}$ in $\mathcal{R}$} 
    \STATE calculate user and item representations;
    \STATE sample user $u$'s interactions $I_u$ in $\mathcal{B}$;
    \STATE divide items $\boldsymbol{I}_u$ to $\boldsymbol{I}_u^{pop}$ and $\boldsymbol{I}_u^{unpop}$ by Eqn.~(\ref{eqa:iu});
    \STATE calculate supervised alignment loss $\mathcal{L}_{sa}$ by Eqn.~(\ref{sa});
    \STATE divide items $\boldsymbol{I}_B$ to $\boldsymbol{I}^{pop}$ and $\boldsymbol{I}^{unpop}$ by Eqn.~(\ref{eqa:ib});
    \STATE calculate re-weighting contrast loss $\mathcal{L}_{cl}$ by Eqn.~(\ref{cl_loss_all});
    \STATE calculate recommendation loss $\mathcal{L}_{rec}$ by Eqn.~(\ref{bprloss});
    \STATE calculate ~\shortname ~ total loss $\mathcal{L}$ by Eqn.~(\ref{overall_equa});
    \STATE update $\Theta \gets \Theta- \eta*\nabla_{\Theta}\mathcal{L}$;
    \IF{early stopping}
    \STATE break;
    \ENDIF
    \ENDFOR
\ENDFOR
\STATE \textbf{return} recommender encoder $f(\Theta)$.
\end{algorithmic}
\end{algorithm}

\section{experiments}
In this section, we evaluate the effectiveness of ~\shortname ~ through extensive experiments, aiming to answer the following questions: 
\begin{itemize}[leftmargin=0.5cm, itemindent=0cm]
    \item \textbf{RQ1}: How does ~\shortname ~ compare to existing debiasing methods?
    \item \textbf{RQ2}: How do different designed components play roles in our proposed ~\shortname?
    \item \textbf{RQ3}: How does ~\shortname ~ alleviate the popularity bias?
    \item \textbf{RQ4}: How do different hyper-parameters affect the ~\shortname ~ recommendation performance?
\end{itemize}
\subsection{Experiments Settings}
\subsubsection{\textbf{Datasets.}} In our experiments, we use three widely public datasets: Amazon-book\footnote{https://jmcauley.ucsd.edu/data/amazon/links.html}, Yelp2018\footnote{https://www.yelp.com/dataset}, and Gowalla\footnote{http://snap.stanford.edu/data/loc-gowalla.html}. 
We retained users and items with a minimum of 10 interactions, consistent with previous works~\cite{Yu2021AreGA, Wei2020ModelAgnosticCR, AutoDebias}.
A detailed description can be found in Appendix A.1.

Note that the traditional dataset splitting methods fail to assess the effectiveness in mitigating popularity bias because the test sets still follow the long-tail distribution~\cite{Wei2020ModelAgnosticCR,zhang2021causal}.
In such cases, the model might perform well during testing even if it heavily relies on popularity for recommendations~\cite{Chen2020BiasAD}.
Hence, the conventional dataset splitting is not appropriate for evaluating whether the model suffers from popularity bias~\cite{Wei2020ModelAgnosticCR}.
To this end, we follow previous works to extract an unbiased dataset where the item distribution in the test set follows a uniform distribution~\cite{zhang2023invariant, zhang2023mitigating, Wei2020ModelAgnosticCR}.
Specifically, we retain a fixed number of interactions for each item in the test set, amounting to approximately 10\% of the entire dataset.
Additionally, to avoid exposing the test distribution, we randomly selected 10\% of interactions from the dataset as the validation set, and the remaining as the training set.

\subsubsection{\textbf{Baselines and Evaluation Metrics.}}

We implement the state-of-the-art LightGCN~\cite{He2020LightGCNSA} to instantiate ~\shortname, aiming to investigate how it alleviates popularity bias.
We compare ~\shortname ~ with several debiased baselines, including re-weighting-based models such as IPS~\cite{Zhu2021PopularityOpportunityBI} and $\gamma$-AdjNorm~\cite{Zhao2022InvestigatingAP}, decorrelation-based models like MACR~\cite{Wei2020ModelAgnosticCR} and InvCF~\cite{zhang2023invariant}, and contrastive learning-based models including Adap-$\tau$~\cite{chen2023adap} and SimGCL~\cite{Yu2021AreGA}. For detailed descriptions of these models, please refer to Appendix A.2.

We utilize three widely used metrics, namely $Recall@K$, $HR@K$, and $NDCG@K$, to evaluate the performance of Top-$K$ recommendation.
$Recall@K$ and $HR@K$ assess the number of target items retrieved in the recommendation results, emphasizing coverage. 
In contrast, $NDCG@K$ evaluates the positions of target items in the ranking list, with a focus on their positions in the list.
Note that we use the full ranking strategy~\cite{zhao2020revisiting}, considering all non-interacted items as candidate items to avoid selection bias during the test stage~\cite{yang2023generative}. 
We repeated each experiment five times with different random seeds and reported the average scores.

\subsubsection{\textbf{Hyper-Parameter Settings.}}
Due to the limited space, more experimental setting details can be found in Appendix A.3.

\subsection{Overall Performance~(RQ1)}
\begin{table*}[t]
    \centering
    \caption{Performance comparison on three public datasets with $K = 20$. The best performance is indicated in bold, while the second-best performance is underlined. The superscripts $*$ indicate $p \leq 0.05$ for the paired t-test of ~\shortname ~ vs. the best baseline (the relative improvements are denoted as Imp.).}
    
    \resizebox{2\columnwidth}{!}{
    \begin{tabular}{c|c|c|c|c|c|c|c|c|c}
    \hline
    \multirow{2}*{\textbf{Model}} & \multicolumn{3}{c|}{\textbf{Yelp2018}} & \multicolumn{3}{c|}{\textbf{Gowalla}} & \multicolumn{3}{c}{\textbf{Amazon-book}}     \\
    \cline{2-10}
     & \textbf{$Recall@20$} &\textbf{ $HR@20$ }& \textbf{$NDCG@20$ }& \textbf{$Recall@20$} &\textbf{ $HR@20$ }& \textbf{$NDCG@20$ }& \textbf{$Recall@20$} &\textbf{ $HR@20$ }& \textbf{$NDCG@20$ }\\

    \hline
    \textbf{MF}&0.0050  &0.0109 &0.0093 &
  0.0343&0.0422 &0.0280&
   0.0370&0.0388&0.0270\\
   \textbf{LightGCN}&0.0048&0.0111 &0.0098&
   0.0380&0.0468&0.0302&
   0.0421&0.0439&0.0304\\
   \hline
    \textbf{IPS}&0.0104&0.0183&0.0158
&0.0562&0.0670&0.0444&
   0.0488&0.0510&0.0365\\
   \textbf{MACR}&0.0402&0.0312&0.0265 &
   0.0908&0.1086&0.0600&
   0.0515&0.0609&0.0487\\
   \textbf{$\gamma$-Adjnorm}&0.0053&0.0088&0.0080
&0.0328&0.0409&0.0267&
   0.0422&0.0450&0.0264\\
   \textbf{InvCF}&0.0444&0.0344&0.0291&
   0.1001&0.1202&0.0662&
   0.0562&0.0665&0.0515\\
   \textbf{Adap-$\tau$}&\underline{0.0450}&0.0497&0.0341
&0.1182&\underline{0.1248}&0.0794&
   \underline{0.0641}&\underline{0.0678}&0.0511\\
   \textbf{SimGCL}&0.0449&\underline{0.0518}&\underline{0.0345}&
   \underline{0.1194}&0.1228&\underline{0.0804}&
   0.0628&0.0648&\underline{0.0525}\\
   \hline
   \textbf{\shortname}&\textbf{0.0494}*&\textbf{0.0574}*&\textbf{0.0375}*&
   \textbf{0.1232}*&\textbf{0.1321}*&\textbf{0.0848}*&
  \textbf{ 0.0701}*&\textbf{0.0724}*&\textbf{0.0556}*\\
   \textbf{Imp.}&\textbf{+9.78  \%}&\textbf{+10.81\%}&\textbf{+8.70\%}&
   \textbf{+3.18\%}&\textbf{+5.85\%}&\textbf{+5.47\%}&
   \textbf{+9.36\%}&\textbf{+6.78\%}&\textbf{5.90\%}\\
    \hline
    \end{tabular}}
    \label{tab:main_table}
\end{table*}

As shown in Table. ~\ref{tab:main_table}, we compare our model with several baselines across three datasets. The best performance for each metric is highlighted in bold, while the second best is underlined. Our model consistently outperforms all compared methods across all metrics in every dataset.

\begin{itemize}[leftmargin=0.5cm, itemindent=0cm]
    
    \item Our proposed model ~\shortname ~ consistently outperforms all baselines and significantly mitigates the popularity bias. Specifically, ~\shortname ~ enhances LightGCN, achieving improvements of 282.65\%, 180.79\%, and 82.89\% in $NDCG@20$ on the Yelp2018, Gowalla, and Amazon-Book datasets, respectively. Compared to the strongest baselines (SimGCL or Adap-$\tau$), ~\shortname ~ delivers better performance. The most significant improvements are observed on Yelp2018, where our model achieves an 8.70\% increase in $Recall@20$, a 10.81\% increase in $HR@20$, and a 30.2\% increase in $NDCG@20$. This improvement can be attributed to our use of popularity-aware supervised alignment to enhance the representation of less popular items and re-weighted contrastive learning to address representation separation from a popularity-centric perspective.

    \item The performance improvements of ~\shortname ~ are smaller on sparser datasets. For example, on the Gowalla dataset, the improvements in $Recall@20$, $HR@20$, and $NDCG@20$ are 3.18\%, 5.85\%, and 5.47\%, respectively. This may be because, in sparser datasets like Gowalla, even popular items are not well-represented due to lower data density. Aligning unpopular items with these poorly represented popular items can introduce noise into the model. Therefore, the benefits of using supervisory signals for unpopular items may be reduced in very sparse environments, leading to smaller performance improvements.
    
    \item Regarding the baselines for mitigating popularity bias, the improvement of $\gamma$-Adjnorm is relatively limited compared to the backbone model (LightGCN) and even performs worse in some cases. This may be because $\gamma$-Adjnorm is specifically designed for traditional data-splitting scenarios, where the test set still follows a long-tail distribution, leading to poor generalization. IPS and MACR mitigate popularity bias by excluding item popularity information. InvCF uses invariant learning to remove popularity information at the representation level, generally performing better than IPS and MACR. This shows the importance of addressing popularity bias at the representation level. Adap-$\tau$ and SimGCL outperform the other baselines, emphasizing the necessary to improve item representation consistency for mitigating popularity bias.

    \item Different metrics across various datasets show varying improvements in model performance. Adapt-$\tau$ performs well on $Recall@20$ and $HR@20$, while SimGCL excels in $NDCG@20$. This suggests that different debiasing methods may need distinct optimization strategies for models. Additionally, we observe varying effects of ~\shortname ~ across different datasets, with performance improvements of 9.76\%, 7.35\%, and 4.83\% on the Yelp2018, Amazon-Book, and Gowalla datasets, respectively. This difference could be due to the sparser nature of the Gowalla dataset. Conversely, our model can directly provide supervisory signals for unpopular items and conduct intra-group optimization, consistently maintaining optimal performance across all metrics on the three datasets. 

\end{itemize}

\subsection{Ablation Study~(RQ2)}

\begin{table*}[t]
    \centering
    \caption{Ablation study of ~\shortname, highlighting the best-performing model on each dataset and metrics in bold. Specifically, ~\shortname-w/o $P$ removes the re-weighting contrastive loss of popular items, ~\shortname-w/o $U$ eliminates the re-weighting contrastive loss of unpopular items, and ~\shortname-w/o $A$ omits the popularity-aware supervised alignment loss.}
    \resizebox{2\columnwidth}{!}{
    \begin{tabular}{c|c|c|c|c|c|c|c|c|c}
    \hline
    \multirow{2}*{\textbf{Model}} & \multicolumn{3}{c|}{\textbf{Yelp2018}} & \multicolumn{3}{c|}{\textbf{Gowalla}} & \multicolumn{3}{c}{\textbf{Amazon-book}}     \\
    \cline{2-10}
     & \textbf{$Recall@20$} &\textbf{ $HR@20$ }& \textbf{$NDCG@20$ }& \textbf{$Recall@20$} &\textbf{ $HR@20$ }& \textbf{$NDCG@20$ }& \textbf{$Recall@20$} &\textbf{ $HR@20$ }& \textbf{$NDCG@20$ }\\

    \hline
   \textbf{SimGCL}&0.0449&0.0518&0.0345&
   0.1194&0.1228&0.0804&
   0.0628&0.0648&0.0525\\

   \hline
   
   \textbf{\shortname-w/o ${P}$}&0.0443&0.0536&0.0340&0.1098&0.1191&0.0750&0.0616&0.0639&0.0458\\
   \textbf{\shortname-w/0 ${U}$}&0.0462&0.0545&0.0358&0.1120&0.1179&0.0752&0.0594&0.0617&0.0464\\
   \textbf{\shortname-w/0 ${A}$}&0.0466&0.0547&0.0360&0.1195&0.1260&0.0815&0.0687&0.0711&0.0536\\
    \hline 
       \textbf{\shortname}&\textbf{0.0494*}&\textbf{0.0574*}&\textbf{0.0375*}&
  \textbf{0.1232*}&\textbf{0.1321*}&\textbf{0.0848*}&
  \textbf{ 0.0701*}&\textbf{0.0724*}&\textbf{0.0556*}\\
   \hline
   
    \end{tabular}}
    \label{tab:ablation_table}
\end{table*}

To better understand the effectiveness of each component in ~\shortname, we conduct ablation studies on three datasets. Table. \ref{tab:ablation_table} presents a comparison between ~\shortname ~ and its variants on recommendation performance. Specifically, ~\shortname-w/o $P$ refers to the variant where the re-weighting contrastive loss of popular items is removed, focusing instead on optimizing the consistency of representations for unpopular items. Similarly, ~\shortname-w/o $U$ denotes the removal of the re-weighting contrastive loss for unpopular items. ~\shortname-w/o $A$ refers to the variant without the popularity-aware supervised alignment loss. It's worth noting that \shortname-w/o $A$ differs from SimGCL in that we split the contrastive loss on the item side, $\mathcal{L}_{cl}^{item}$, into two distinct losses: $\mathcal{L}^{pop}_{cl}$ and $\mathcal{L}^{unpop}_{cl}$. This approach allows us to separately address the consistency of popular and unpopular item representations, thereby providing a more detailed analysis of the impact of each component on the overall performance.

From Table.~\ref{tab:ablation_table}, we observe that ~\shortname-w/o $A$ outperforms SimGCL in most cases. This validates that re-weighting the importance of popular and unpopular items can effectively improve the model's performance in alleviating popularity bias. It also demonstrates the effectiveness of using supervision signals from popular items to enhance the representations of unpopular items, providing more opportunities for future research on mitigating popularity bias. Moreover, compared with ~\shortname-w/o $U$, ~\shortname-w/o $P$ results in much worse performance. This confirms the importance of re-weighting popular items in contrastive learning for mitigating popularity bias. Finally, ~\shortname ~ consistently outperforms the three variants, demonstrating the effectiveness of combining supervised alignment and re-weighting contrastive learning. Based on the above analysis, we conclude that leveraging supervisory signals from popular item representations can better optimize representations for unpopular items, and re-weighting contrastive learning allows the model to focus on more informative or critical samples, thereby improving overall performance. All the proposed modules significantly contribute to alleviating popularity bias.

\subsection{Debias Ability~(RQ3)}

\begin{figure}[t]
    \centering
    \subfloat{\includegraphics[width =0.5\linewidth]{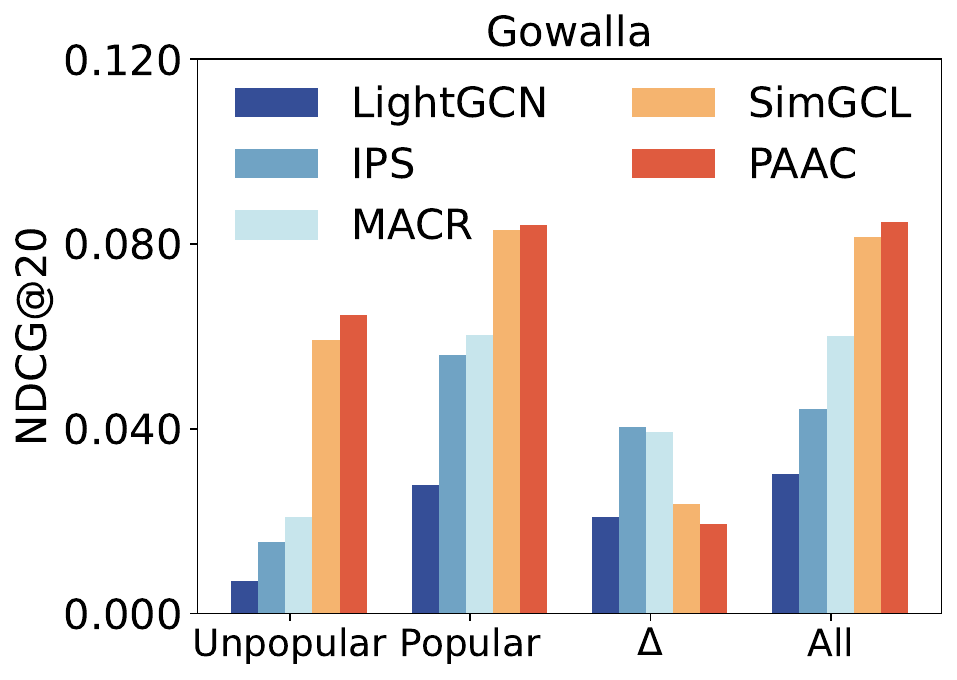}\label{fig:NDCG20_gowalla}}
    \hfill
    \subfloat{\includegraphics[width =0.5\linewidth]{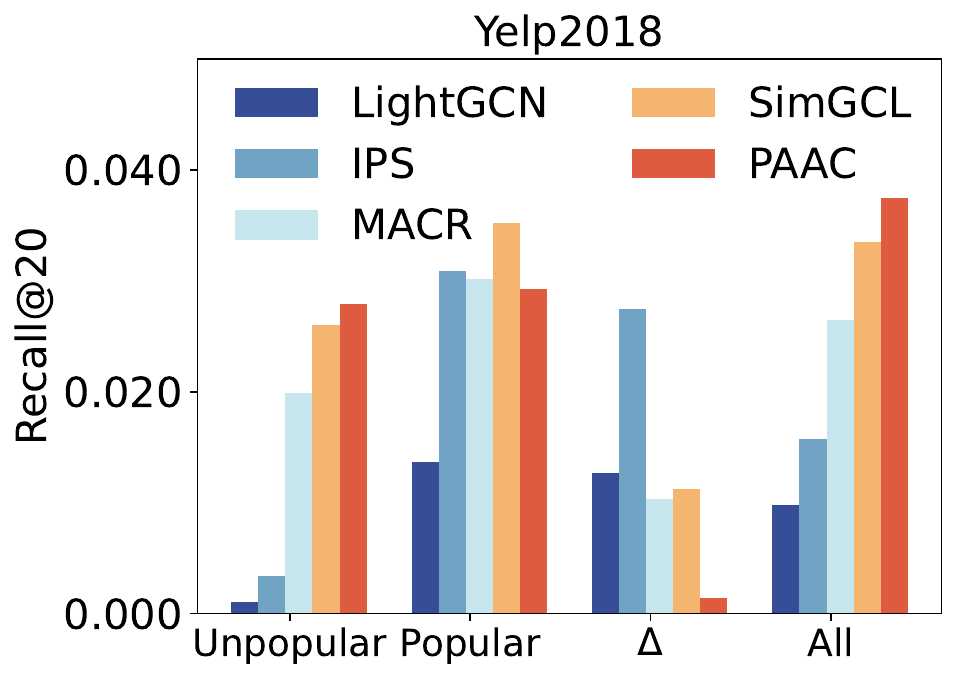}\label{fig:NDCG20_yelp}}
    \caption{Performance comparison over different item popularity groups. In particular, $\Delta$ indicates the accuracy gap between different groups.}
    \label{fig:NDCG=recall}
\end{figure}

To further verify the effectiveness of ~\shortname ~ in alleviating popularity bias, we conduct a comprehensive analysis focusing on the recommendation performance across different popularity item groups. Specifically, 20\% of the most popular items are labeled ‘Popular’, and the rest are labeled ‘Unpopular’. As shown in Figure. ~\ref{fig:NDCG=recall}, we compare the performance of ~\shortname ~ with LightGCN, IPS, MACR, and SimGCL using the $NDCG@20$ metric across different popularity groups. We use $\Delta$ to denote the accuracy gap between the two groups. From Figure. ~\ref{fig:NDCG=recall}, we draw the following conclusions:

\begin{itemize}[leftmargin=0.5cm, itemindent=0cm]

    \item Our proposed ~\shortname ~ significantly enhances the recommendation performance for unpopular items. Specifically, we observe an improvement of 8.94\% and 7.30\% in $NDCG@20$ relative to SimGCL on the Gowalla and Yelp2018 datasets, respectively. This improvement is due to the popularity-aware supervised alignment method, which uses supervisory signals from popular items to improve the representations of unpopular items.
    
    \item \shortname ~ has successfully narrowed the accuracy gap between different item groups. Specifically,  ~\shortname ~ achieved the smallest gap, reducing the $NDCG@20$ accuracy gap by 34.18\% and 87.50\% on the Gowalla and Yelp2018 datasets, respectively. This indicates that our method treats items from different groups fairly, effectively alleviating the impact of popularity bias. This success can be attributed to our re-weighted contrast module, which addresses representation separation from a popularity-centric perspective, resulting in more consistent recommendation results across different groups.

    \item Improving the performance of unpopular items is crucial for enhancing overall model performance. Specially, on the Yelp2018 dataset, ~ \shortname ~ shows reduced accuracy in recommending popular items, with a notable decrease of 20.14\% compared to SimGCL. However, despite this decrease, the overall recommendation accuracy surpasses that of SimGCL by 11.94\%, primarily due to a 6.81\% improvement in recommending unpopular items. This improvement highlights the importance of better recommendations for unpopular items and emphasizes their crucial role in enhancing overall model performance.

\end{itemize}

Due to space limitations, we demonstrate the debiasing capabilities of our model from more dimensions in Appendix A.4, such as conventional test sets and representation separation analysis.

\subsection{Hyperparameter Sensitivities~(RQ4)}

In this section, we analyze the impact of hyperparameters in ~\shortname. Firstly, we investigate the influence of $\lambda_1$ and $\lambda_2$, which respectively control the impact of the popularity-aware supervised alignment and re-weighting contrast loss. Additionally, in the re-weighting contrastive loss, we introduce two hyperparameters, $\gamma$ and $\beta$, to control the re-weighting of different popularity items as positive and negative samples. Finally, we explore the impact of the grouping ratio $x$ on the model's performance.

\subsubsection{\textbf{Effect of $\lambda_{1}$  and $\lambda_{2}$.}}

\begin{figure}
    \vspace{-0.2cm}
    \centering
    \subfloat{\includegraphics[width =0.5\linewidth]{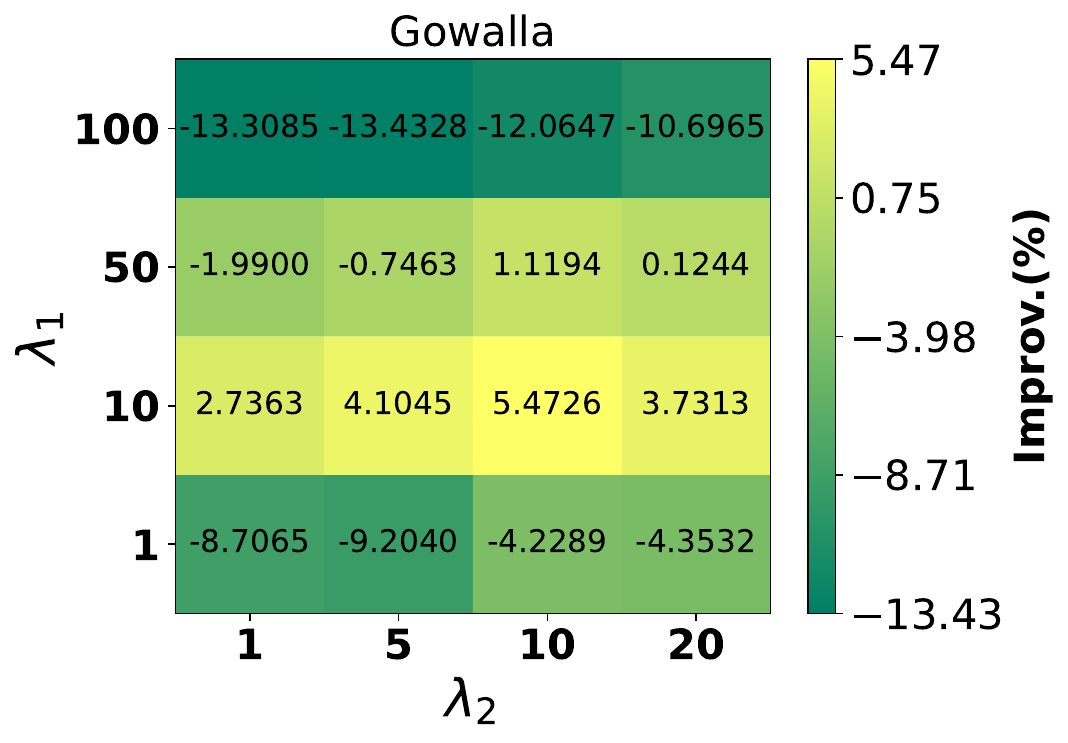}\label{fig:ndcg_gowalla_hyper}}
    \hfill
    \subfloat{\includegraphics[width =0.5\linewidth]{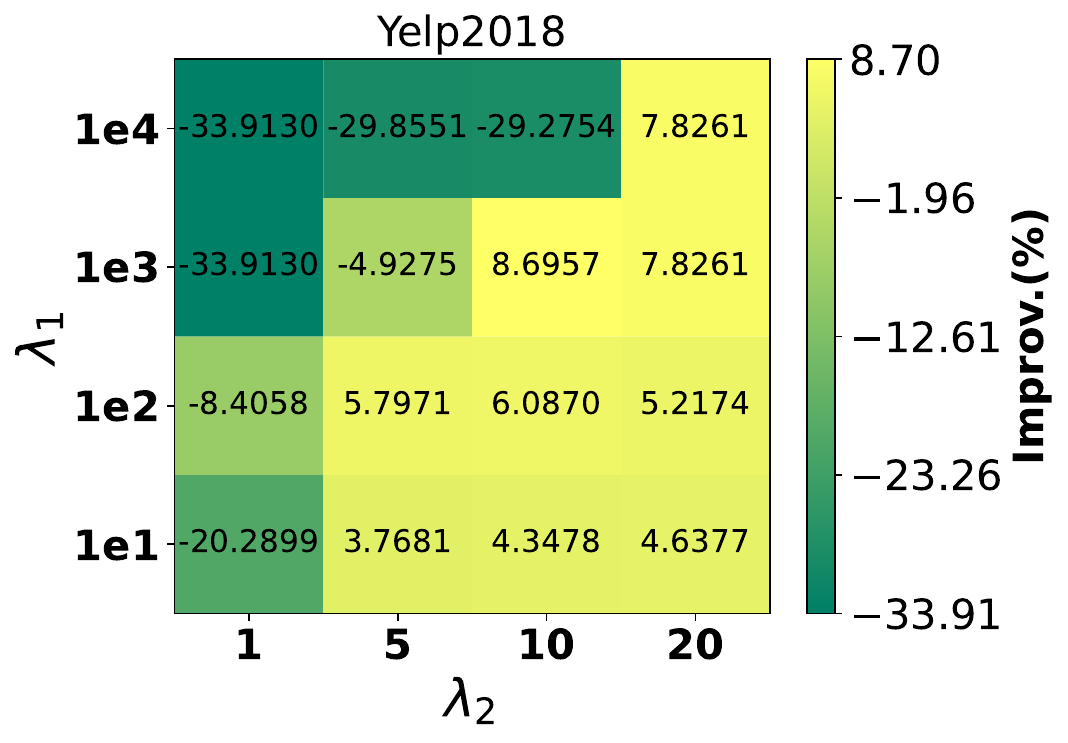}\label{fig:ndcg_yelp_hyper}}
    \caption{Performance comparison w.r.t. $\lambda_{1}$ and $\lambda_{2}$ on the Yelp2018 and Gowalla dataset in $NDCG@20$. The values indicate the percentage improvement relative to the best baseline.}
    \label{fig:yelp2018-hyper2}
    \vspace{-0.3cm}
\end{figure}
As formulated in Eq. (\ref{overall_equa}), $\lambda_{1}$ controls the extent of providing additional supervisory signals for unpopular items, while $\lambda_{2}$ controls the extent of optimizing representation consistency. Figure. ~\ref{fig:yelp2018-hyper2} illustrates how the relative performance to the best baseline $NDCG@20$ varies with $\lambda_1$ and $\lambda_2$ on the Yelp2018 and Gowalla datasets.
\textbf{Horizontally}, with the increase in $\lambda_{2}$, the performance initially increases and then decreases. This indicates that appropriate re-weighting contrastive loss effectively enhances the consistency of representation distributions, mitigating popularity bias. However, overly strong contrastive loss may lead the model to neglect recommendation accuracy.
\textbf{Vertically}, as $\lambda_{1}$ increases, the performance also initially increases and then decreases. This suggests that suitable alignment can provide beneficial supervisory signals for unpopular items, while too strong an alignment may introduce more noise from popular items to unpopular ones, thereby impacting recommendation performance.

\subsubsection{\textbf{Effect of re-weighting coefficient $\gamma$ and $\beta$.}}
\begin{figure}
    \vspace{-0.2cm}
    \centering
    \subfloat{\includegraphics[width =0.5\linewidth]{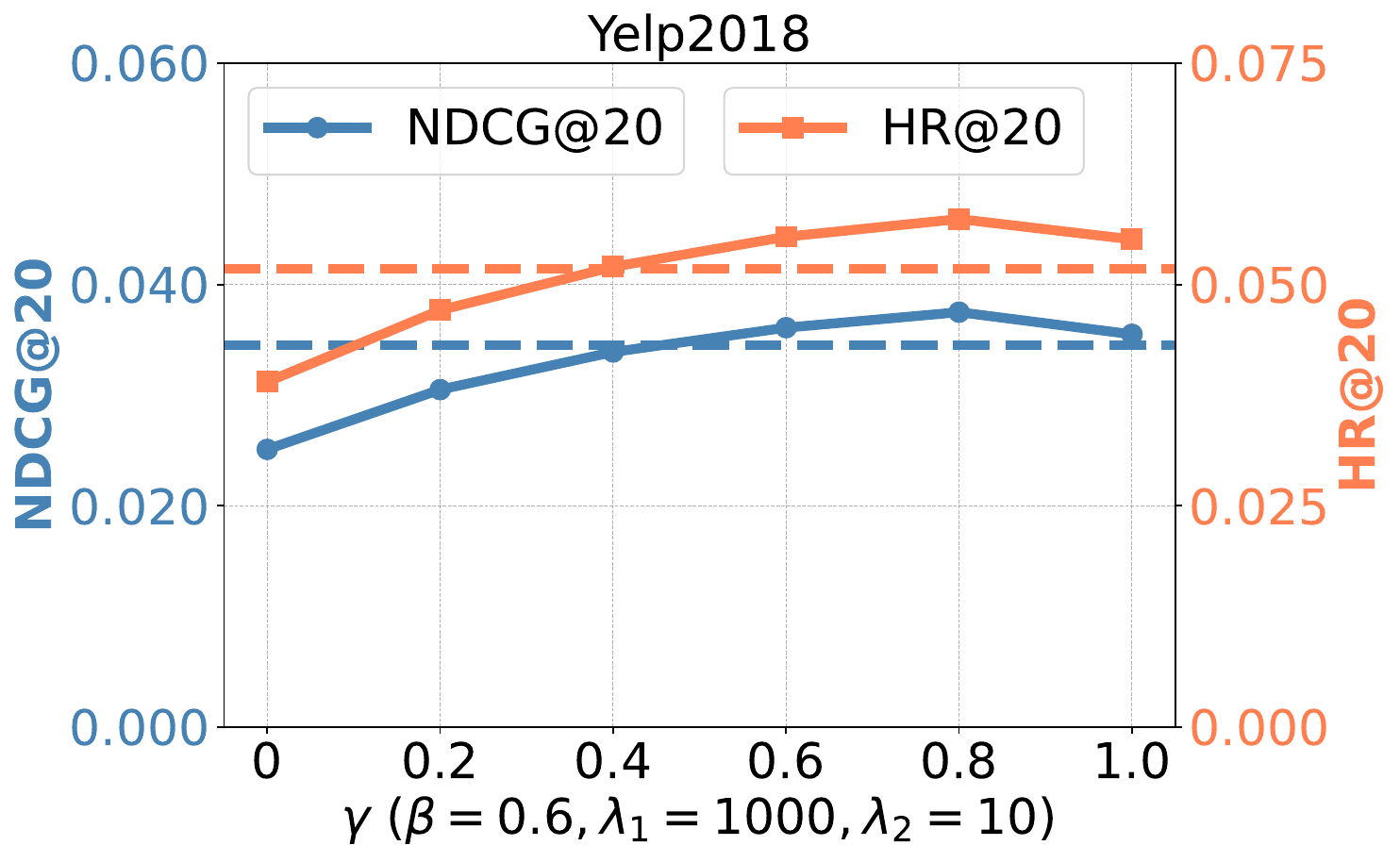}\label{fig:yelp_gamma}}
    \hfill
    \subfloat{\includegraphics[width =0.5\linewidth]{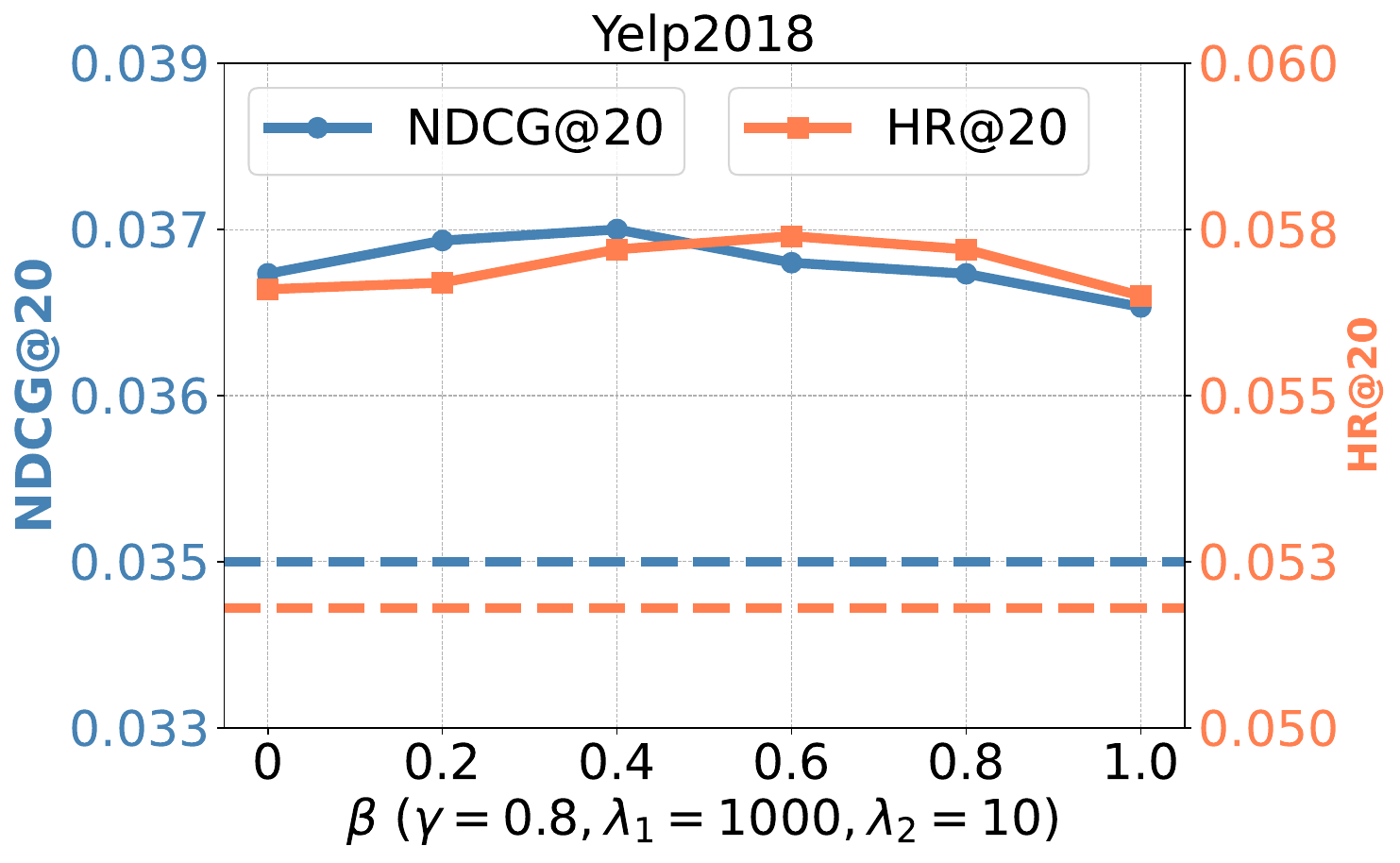}\label{fig:yelp_lambda}}
    \hfill
    \subfloat{\includegraphics[width =0.5\linewidth]{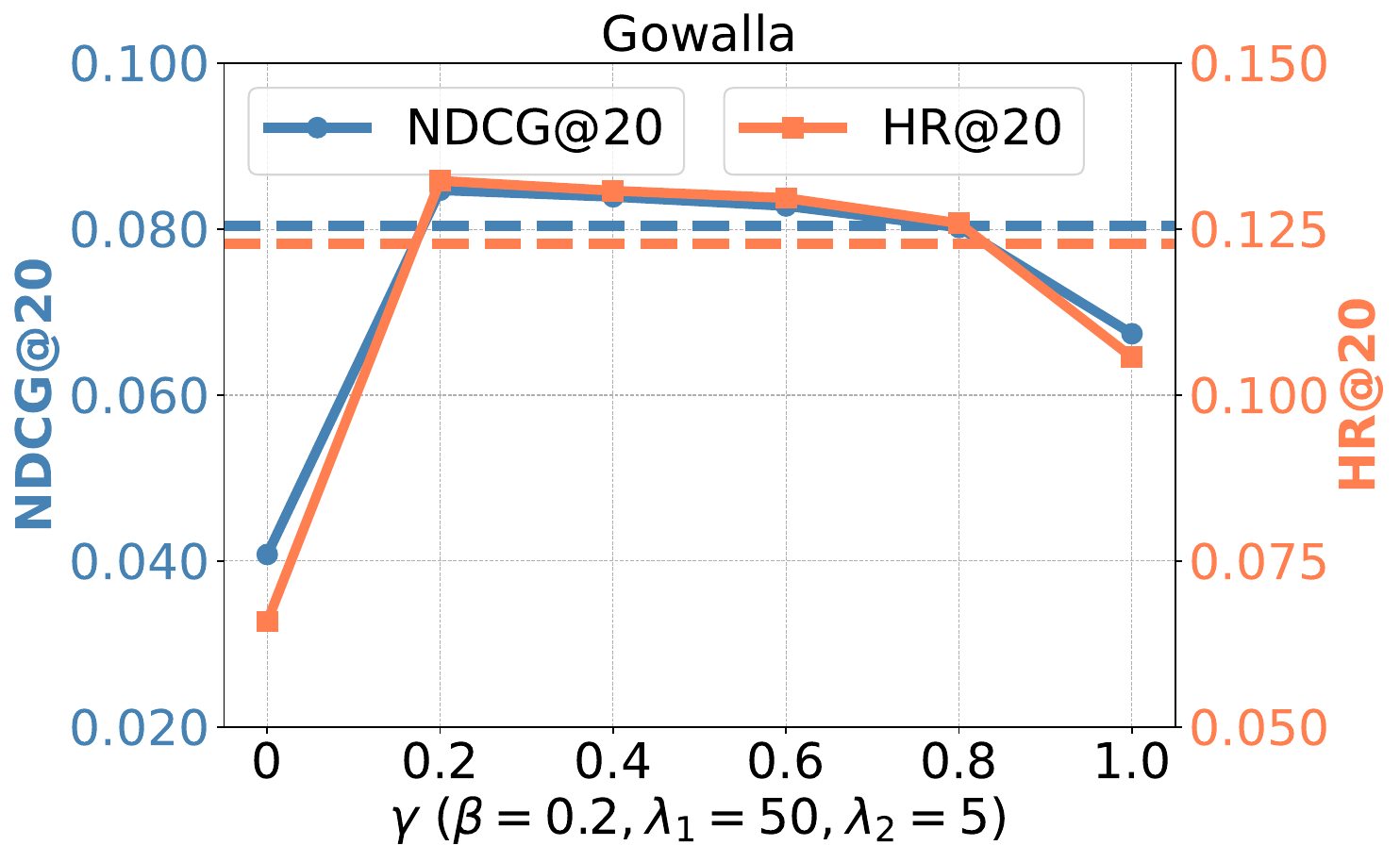}\label{fig:gowalla_gamma}}
    \subfloat{\includegraphics[width =0.5\linewidth]{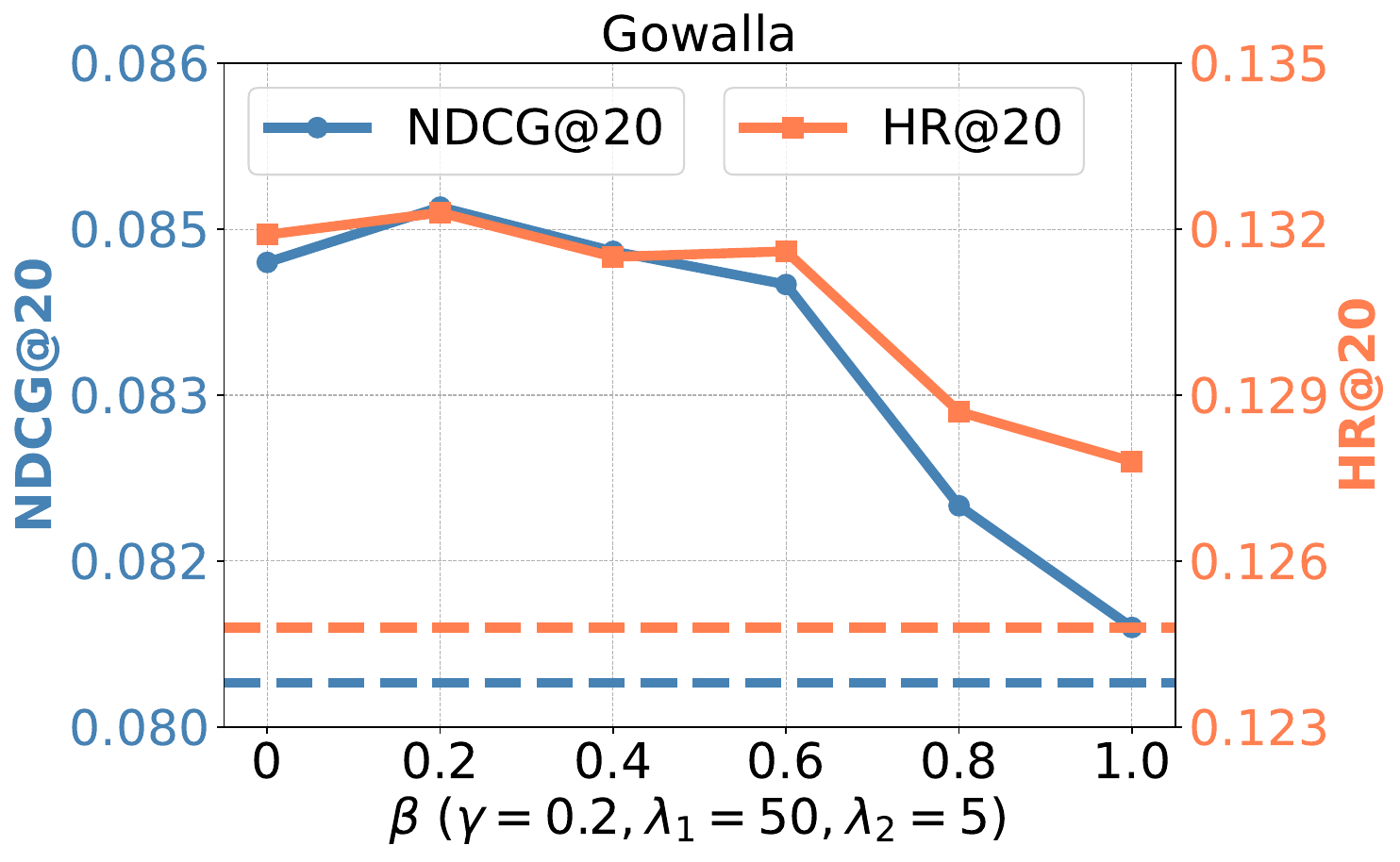}\label{fig:gowalla_lambda}}
    \caption{Performance comparison w.r.t. different  $\gamma$ and $\beta$. The top shows the $NDCG@20$ and $HR@20$ results on Yelp2018 and the bottom shows the results on Gowalla. The horizontal line represents the best results already achieved in the baseline.}
    \label{fig:gamma and beta}
    \vspace{-0.2cm}
\end{figure}

To mitigate representation separation due to imbalanced positive and negative sampling, we introduce two hyperparameters into the contrastive loss. Specifically, $\gamma$ controls the weight difference between positive samples from popular and unpopular items, while $\beta$ controls the influence of different popularity items as negative samples.

In our experiments, while keeping other hyperparameters constant, we search $\gamma$ and $\beta$ within the range \{0, 0.2, 0.4, 0.6, 0.8, 1\}. Figure.~\ref{fig:gamma and beta} illustrates how performance changes when varying $\gamma$ and $\beta$ on two datasets, with horizontal lines representing the best baseline. As $\gamma$ and $\beta$ increase, performance initially improves and then declines. The optimal hyperparameters for the Yelp2018 and Gowalla datasets are $\gamma=0.8$, $\beta=0.6$ and $\gamma=0.2$, $\beta=0.2$, respectively.
This may be attributed to the characteristics of the datasets. The Yelp2018 dataset, with a higher average interaction frequency per item, benefits more from a higher weight $\gamma$ for popular items as positive samples. Conversely, the Gowalla dataset, being relatively sparse, prefers a smaller $\gamma$. This indicates the importance of considering dataset characteristics when adjusting the contributions of popular and unpopular items to the model.

Notably, $\gamma$ and $\beta$ are not highly sensitive within the range \([0,1]\), performing well across a broad spectrum. Figure.~\ref{fig:gamma and beta} shows that performance exceeds the baseline regardless of $\beta$ values when other parameters are optimal. Additionally, $\gamma$ values from \([0.4, 1.0]\) on the Yelp2018 dataset and \([0.2, 0.8]\) on the Gowalla dataset surpass the baseline, indicating less need for precise tuning. Thus, $\gamma$ and $\beta$ achieve optimal performance without meticulous adjustments, focusing on weight coefficients to maintain model efficacy.

\subsubsection{\textbf{Effect of grouping ratio $x$.}}

\begin{table}
\centering
\vspace{-0.4cm}
\caption{Performance comparison across varying popular item ratios $x$ on metrics.}
\vspace{-0.2cm}
\label{fig:radio}
\resizebox{1\columnwidth}{!}{
\begin{tabular}{|l|l|l|l|l|l|l|}
\hline
        \multirow{2}*{\textbf{Ratio}} & \multicolumn{3}{c|}{\textbf{Yelp2018}}& \multicolumn{3}{c|}{\textbf{Gowalla}} \\
        \cline{2-7}
        &$Recall@20$ & $HR@20$ & $NDCG@20$ & $Recall@20$ & $HR@20$ & $NDCG@20$  \\ \hline
                       
\textbf{20\% }                  & 0.0467    & 0.0555 & 0.0361  & 0.1232    & 0.1319 & 0.0845  \\ \hline
\textbf{40\%  }                 & 0.0505    & 0.0581 & 0.0378  & 0.1239    & 0.1325 & 0.0848 \\ \hline
\textbf{50\% }                  & 0.0494    & 0.0574 & 0.0375  & 0.1232    & 0.1321 & 0.0848  \\ \hline
\textbf{60\% }                  & 0.0492    & 0.0569 & 0.0370  & 0.1225    & 0.1314 & 0.0843 \\ \hline
\textbf{80\% }                  & 0.0467    & 0.0545 & 0.0350  & 0.1176    & 0.1270 & 0.0818  \\ \hline
\end{tabular}}
\vspace{-0.6cm}
\end{table}

To investigate the impact of different grouping ratios on recommendation performance, we developed a flexible classification method for items within each mini-batch based on their popularity. Instead of adopting a fixed global threshold, which tends to overrepresent popular items in some mini-batches, our approach dynamically divides items in each mini-batch into popular and unpopular categories. Specifically, the top $x\%$ of items are classified as popular and the remaining $(100-x)\%$ as unpopular, with $x$ varying. This strategy prevents the overrepresentation typical in fixed distribution models, which could skew the learning process and degrade performance.
To quantify the effects of these varying ratios, we examined various division ratios for popular items, including 20\%, 40\%, 60\%, and 80\%, as shown in Table.~\ref{fig:radio}. The preliminary results indicate that both extremely low and high ratios negatively affect model performance, thereby underscoring the superiority of our dynamic data partitioning approach. Moreover, within the 40\%-60\% range, our model's performance remained consistently robust, further validating the effectiveness of ~\shortname.

\section{Related Work}
\subsection{Popularity Bias in Recommendation}
Popularity bias is a common issue in recommender systems where unpopular items in the training dataset are rarely recommended~\cite{Wei2020ModelAgnosticCR,DecRS}. 
Many methods~\cite{Chen2020BiasAD, wang2023survey,wang2023unbiased,wang2024intersectional,cai2024mitigating} have been proposed to analyze and reduce performance differences between popular and unpopular items. These methods can be broadly categorized into three types.

\begin{itemize}[leftmargin=0cm, itemindent=0.3cm]
    \item  \textbf{Re-weighting-based methods} aim to increase the training weight or scores for unpopular items, redirecting focus away from popular items during training or prediction~\cite{Zhao2022InvestigatingAP, Zhu2021PopularityOpportunityBI, AutoDebias}. For instance, IPS~\cite{Zhu2021PopularityOpportunityBI} adds compensation to unpopular items and adjusts the prediction of the user-item preference matrix, resulting in higher preference scores and improving rankings for unpopular items. $\gamma$-AdjNorm ~\cite{Zhao2022InvestigatingAP} enhances the focus on unpopular items by controlling the normalization strength during the neighborhood aggregation process in GCN-based models.
    
    \item \textbf{Decorrelation-based methods} aim to effectively remove the correlations between item representations (or prediction scores) and popularity\cite{Wei2020ModelAgnosticCR, wu2021learning, bonner2018causal,shaofair,zhang2021causal,wang2022causal}. For instance, MACR~\cite{Wei2020ModelAgnosticCR} uses counterfactual reasoning to eliminate the direct impact of popularity on item outcomes. In contrast, InvCF~\cite{zhang2023invariant} operates on the principle that item representations remain invariant to changes in popularity semantics, filtering out unstable or outdated popularity characteristics to learn unbiased representations.
    
    \item \textbf{Contrastive-learning-based methods} aim to achieve overall uniformity in item representations using InfoNCE ~\cite{jaiswal2020survey,yang2021enhanced}, preserving more inherent characteristics of items to mitigate popularity bias~\cite{wang2020understanding, Yu2021AreGA}. This approach has been demonstrated as a state-of-the-art method for alleviating popularity bias. It employs data augmentation techniques such as graph augmentation or feature augmentation to generate different views, maximizing positive pair consistency and minimizing negative pair consistency to promote more uniform representations~\cite{Yao2020SelfsupervisedLF}. Specifically, Adap-$\tau$\cite{chen2023adap} adjusts user/item embeddings to specific values, while SimGCL\cite{Yu2021AreGA} integrates InfoNCE loss to enhance representation uniformity and alleviate popularity bias.
\end{itemize}

\subsection{Representation Learning for CF}
Representation learning is crucial in recommendation systems, especially in modern collaborative filtering (CF) techniques. It creates personalized embeddings that capture user preferences and item characteristics~\cite{Rendle2009BPRBP, citationsurveylekey, jaiswal2020survey,ye2022towards,wu2023causality}. The quality of these representations critically determines a recommender system's effectiveness by precisely capturing the interplay between user interests and item features~\cite{Wang2022TowardsRA, Yu2021AreGA,he2024double}. Recent studies emphasize two fundamental principles in representation learning: alignment and uniformity~\cite{wang2020understanding, Wang2022TowardsRA, sun2023neighborhood}. The alignment principle ensures that embeddings of similar or related items (or users) are closely clustered together, improving the system's ability to recommend items that align with a user's interests~\cite{wang2020understanding}. This principle is crucial when accurately reflecting user preferences through corresponding item characteristics~\cite{Wang2022TowardsRA}. Conversely, the uniformity principle ensures a balanced distribution of all embeddings across the representation space~\cite{Yu2021AreGA, Wu2020SelfsupervisedGL}. This approach prevents the over-concentration of embeddings in specific areas, enhancing recommendation diversity and improving generalization to unseen data~\cite{Yu2021AreGA}.

In this work, we focus on aligning the representations of popular and unpopular items interacted with by the same user and re-weighting uniformity to mitigate representation separation. Our model ~\shortname ~uniquely addresses popularity bias by combining group alignment and contrastive learning, a first in the field. Unlike previous works that align positive user-item pairs or contrastive pairs, ~\shortname ~ directly aligns popular and unpopular items, leveraging the rich information of popular items to enhance the representations of unpopular items and reduce overfitting. Additionally, we introduce targeted re-weighting from a popularity-centric perspective to achieve a more balanced representation. 

\section{Conclusion}
In this work, we analyzed popularity bias and proposed ~\shortname ~ to mitigate popularity bias. We assumed that items interacted with by the same user share similar characteristics and used this observation to align representations of both popular and unpopular items through a popularity-aware supervised alignment approach. This provided more supervisory information for unpopular items. Note that our hypothesis of aligning and grouping items based on user-specific preferences offers a novel alignment perspective. Additionally, we addressed the issue of representation separation in current CL-based models by introducing two hyper-parameters to control the weights of items with different popularity levels as positive and negative samples. This approach optimized representation consistency and effectively alleviated separation. Our method, ~\shortname, was validated on three public datasets, proving its rationale and effectiveness. 

In the future, we will explore deeper alignment and contrast adjustments tailored to specific tasks to further mitigate popularity bias. We aim to investigate the synergies between alignment and contrast and extend our approach to address other biases in recommendation systems.


\begin{acks}
This work was supported in part by grants from the National Key Research and Development Program of China (Grant No. 2021ZD0111802), the National Natural Science Foundation of China (Grant No. 72188101, U21B2026), the Fundamental Research Funds for the Central Universities, and Quan Cheng Laboratory (Grant No. QCLZD202301).
\end{acks}

\bibliographystyle{ACM-Reference-Format}
\bibliography{sample-base}

\appendix
\section{APPENDIX}
\subsection{Datasets}
We conducted experiments on three public datasets, with the processed dataset statistics summarized in Table. ~\ref{tab:stats}. Additionally, to analyze the imbalance in item distribution, we recorded the number of interactions for the most and least popular items and calculated the average number of interactions per item. We also used the Gini coefficient to reflect the disparity in item popularity distribution~\cite{park2023toward}. The results are presented in Table. ~\ref{tab:gini}.

\textbf{Amazon-Book.} Amazon is a frequently utilized dataset for item recommendations. From this collection, we specifically selected the Amazon-Book dataset.

\textbf{Yelp2018.} This dataset is sourced from the 2018 edition of the Yelp Challenge, where local businesses such as restaurants and bars are considered items.

\textbf{Gowalla.} This is a check-in dataset from Gowalla, that contains user location data shared through check-ins.

\begin{table}[h]
    \centering
    \caption{The statistics of three datasets.}
    \resizebox{1\columnwidth}{!}{
    \begin{tabular}{ccccl}
    \toprule
   \textbf{ Datasets}&\textbf{\#Users}&\textbf{\#Itmes}&\textbf{\#Interactions}&\textbf{Density}\\
    \midrule
    \textbf{Amazon-Book}&52,643&91,599&2,984,108&0.0619\%\\
    \textbf{Yelp2018}&31,668&38,048&1,561,406&0.1300\%\\
    \textbf{Gowalla}&29,858&40,981&1,027,370&0.0840\%\\
    \bottomrule
    \end{tabular}}
    \label{tab:stats}
\end{table}

\begin{table}[h]
    \centering
    \caption{The analysis of item popularity distribution.}
    \resizebox{1\columnwidth}{!}{
    \begin{tabular}{ccccc}
    \toprule
   \textbf{ Datasets}&\textbf{\#Max}&\textbf{\#Min}&\textbf{\#Average}&\textbf{GINI}\\
    \midrule
    \textbf{Amazon-Book}& 1902&5& 27.58&0.55 \\
    \textbf{Yelp2018}&744 &3 &18.28 &0.58 \\
    \textbf{Gowalla}&2305 & 5& 20.07&0.55 \\
    \bottomrule
    \end{tabular}}
    \label{tab:gini}
\end{table}

\subsection{Baselines}
We compare ~\shortname ~ with several debiased baselines, including re-weighting-based models such as IPS and $\gamma$-AdjNorm, decorrelation-based models like MACR and InvCF, and contrastive learning-based models including Adap-$\tau$ and SimGCL.
\begin{itemize}[leftmargin=0.5cm, itemindent=0cm]
    \item \textbf{BPRMF~\cite{Rendle2009BPRBP}} is a traditional CF-based model that maps user and item IDs into a representation space via matrix factorization. It optimizes using Bayesian Personalized Ranking (BPR) loss;
    \item \textbf{LightGCN~\cite{He2020LightGCNSA}} is a state-of-the-art CF-based model that effectively captures high-order collaborative signals between users and items by linearly propagating embeddings on the user-item interaction graph through multiple layers;
    \item \textbf{IPS~\cite{Zhu2021PopularityOpportunityBI}} adjusts the weight of each user-item interaction according to item popularity, aiming to mitigate popularity bias;
    \item \textbf{MACR~\cite{Wei2020ModelAgnosticCR}} estimates and eliminates the direct influence of item popularity on prediction scores, using counterfactual inference to mitigate popularity bias;
    \item \textbf{$\gamma$-AdjNorm~\cite{Zhao2022InvestigatingAP}} modifies the hyper-parameter $\gamma$ to preferentially treat unpopular items in graph aggregation, resulting in non-symmetric aggregation;
    \item \textbf{InvCF~\cite{zhang2023invariant}} learns unbiased preference representations that remain stable regardless of item popularity, simultaneously removing unstable and outdated characteristics;
    \item \textbf{Adap-$\tau$~\cite{chen2023adap}} dynamically standardizes user and item embeddings to specific values to reduce popularity bias;
    \item \textbf{SimGCL~\cite{Yu2021AreGA}} aims to achieve a more uniform distribution of representations by incorporating the InfoNCE loss, which helps in mitigating popularity bias.
\end{itemize}

\subsection{Hyper-Parameter Settings} 

We initialize parameters using the Xavier initializer~\cite{defferrard2016convolutional} and use the Adam optimizer~\cite{kingma2014adam} with a learning rate of 0.001.
The embedding size is fixed at 64.
For all datasets, we set the batch size is 2048, and the $L_2$ regularization coefficient $\lambda_3$ is 0.0001.
For our model, we fine-tune the popularity-aware supervised alignment coefficient $\lambda_1$ in \{1, 5, 10, 50, 100, 300, 400, 500, 100\}, the popularity-aware contrastive regularization coefficient $\lambda_2$ in \{0.1, 1, 5, 10, 20\}, and the re-weighting hyperparameters $\gamma$ and $\beta$ in \{0, 0.2, 0.4, 0.5, 0.6, 0.8, 1.0\}.
Additionally, we set the grouping ratio  x=50. This means we dynamically divided items into popular and unpopular groups within each mini-batch based on their popularity, assigning the top 50\% as popular items and the bottom 50\% as unpopular items. This approach ensures equal representation of both groups in our contrastive learning and allows items to be adaptively classified based on the batch's current composition.
Moreover, we carefully search the hyper-parameters for all baselines to ensure fair comparisons.

\subsection{Debias Ability}
\subsubsection{\textbf{Conventional Test.}} As mentioned in Section 4.1.1, traditional evaluation methods fail to accurately measure a model's ability to mitigate popularity bias, as test sets often exhibit a long-tail distribution. This can misleadingly suggest high performance for models that favor popular items. To address this significant issue, we utilize an unbiased dataset for evaluation, ensuring a uniform item distribution in the test set, following established precedents. Despite this, traditional performance metrics remain essential for effectively assessing model performance. Therefore, we also examine ~\shortname's results under conventional experimental settings. These results, detailed in Table.~\ref{tab:iid}, demonstrate that our method competes closely with the best baselines, affirming its efficacy under standard evaluation conditions.

\begin{table*}
\centering
\caption{Performance comparison on conventional test.}
\label{tab:iid}
\begin{tabular}{|l|l|l|l|l|l|l|}
\hline
        \multirow{2}*{\textbf{Ratio}} & \multicolumn{3}{c|}{\textbf{Yelp2018}}& \multicolumn{3}{c|}{\textbf{Gowalla}} \\
        \cline{2-7}
        &$Recall@20$ & $HR@20$ & $NDCG@20$ & $Recall@20$ & $HR@20$ & $NDCG@20$  \\ \hline
\textbf{LightGCN}                      & 0.0591    & 0.0614 & 0.0483       & 0.1637    & 0.1672 & 0.1381 \\ \hline
\textbf{Adapt-$\tau$} & 0.0724    & 0.0753 & 0.0603       & 0.1889    & 0.1930 & 0.1584\\ \hline
\textbf{SimGCL  }                      & 0.0720    & 0.0748 & 0.0594       & 0.1817    & 0.1858 & 0.1526  \\ \hline
\textbf{\shortname}   & 0.0722    & 0.0755 & 0.0602       & 0.1890    & 0.1928 & 0.1585  \\ \hline
\end{tabular}
\end{table*}

\subsubsection{\textbf{Impact on Embedding Separation.}}
To evaluate the impact of ~\shortname ~ on embedding separation, we conduct detailed quantitative and qualitative analyses from various perspectives.

\begin{table*}[t]
\centering
\caption{Comparative Analysis of Maximum Mean Discrepancy (MMD) and Cosine Similarity (CS) Metrics. This table illustrates the effectiveness of the PAAC method in reducing distribution disparities between popular and unpopular item groups. The lower MMD and CS values indicate improved embedding separation, demonstrating the efficacy of the PAAC method.}
\label{tab:quantitative}
\begin{tabular}{|c|c|c|c|c|c|c|}
        \hline
        \textbf{Metrics} & \multicolumn{3}{c|}{\textbf{MMD}$\downarrow$} & \multicolumn{3}{c|}{\textbf{CS}$\downarrow$} \\
        \hline
        \textbf{Model} & LightGCN & SimGCL & \shortname & LightGCN & SimGCL & \shortname \\
        \hline
\textbf{Gowalla}                & 0.001233 & 0.001139 & \textbf{0.000978} & 0.010066 & 0.001202 & \textbf{0.000585}  \\ \hline   
\textbf{Yelp2018}              & 0.000808 & 0.000718 & \textbf{0.000523 }& 0.018836 & 0.001154  & \textbf{0.000400} \\ \hline   
\end{tabular}
\end{table*}

\begin{figure*}[t]
    \centering
    \subfloat{\includegraphics[width =0.255\linewidth]{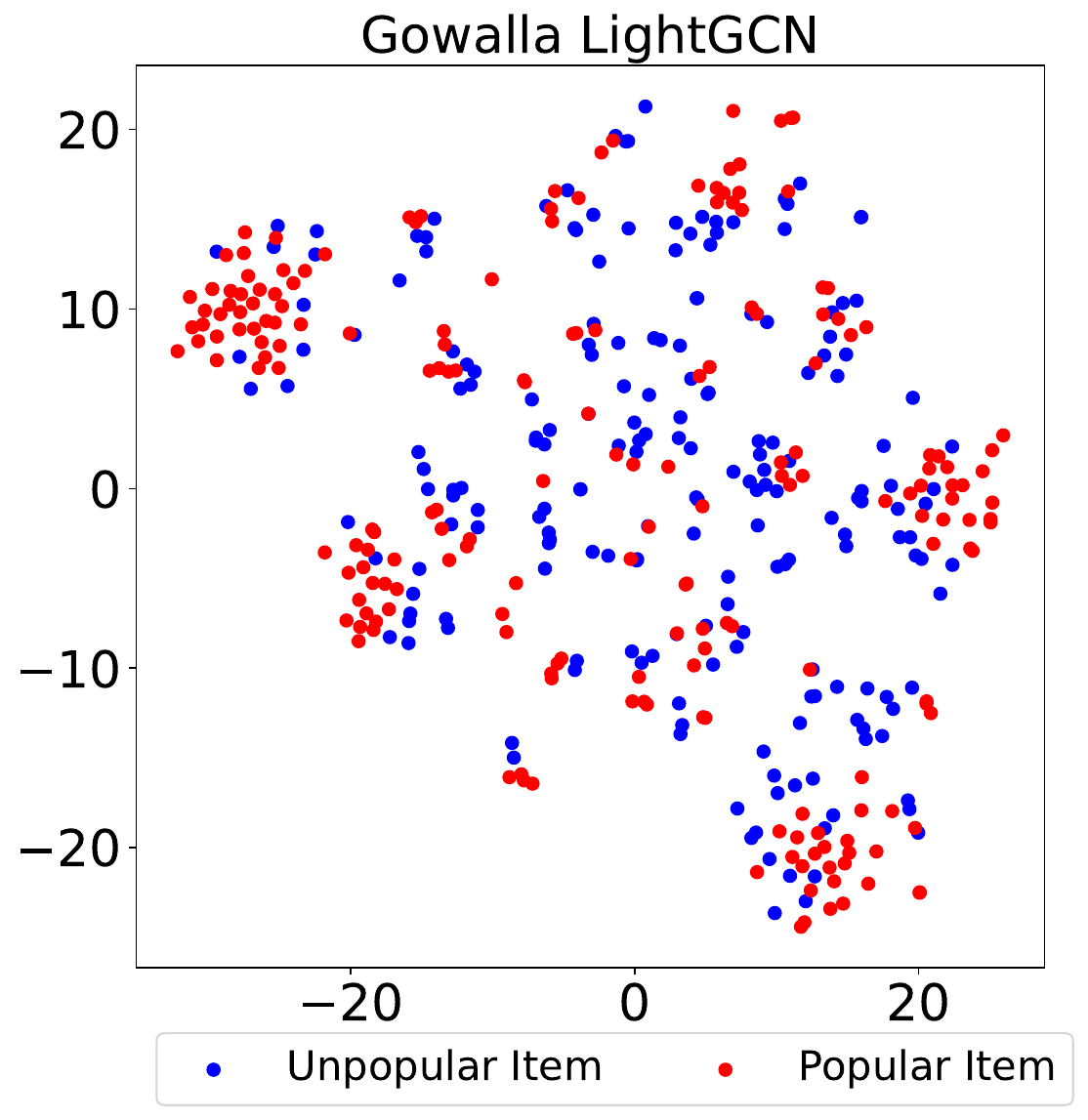}\label{fig:a_lgcn}}
    \hfill
    \subfloat{\includegraphics[width =0.245\linewidth]{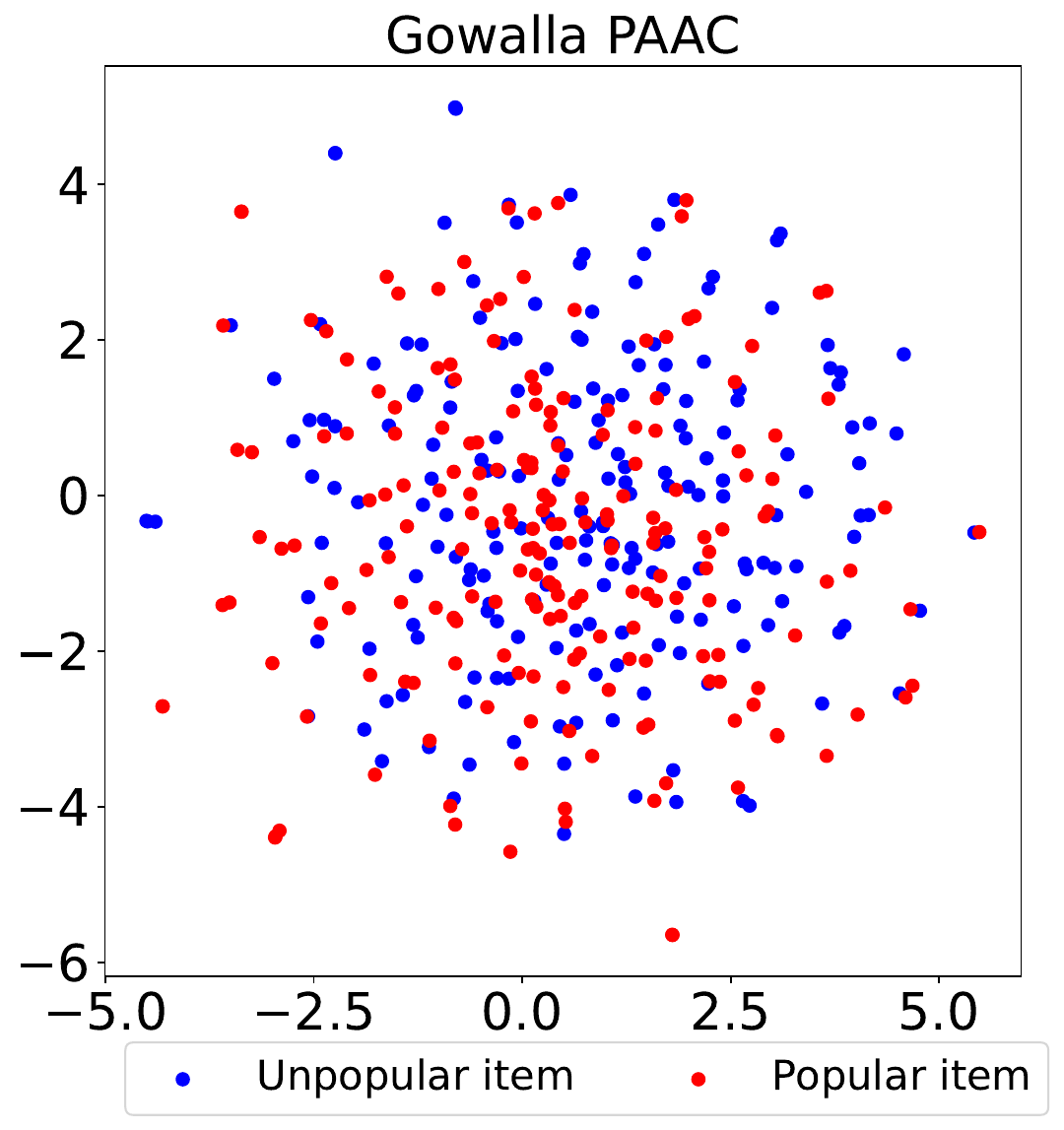}\label{fig:a_ours}}
    \hfill
    \subfloat{\includegraphics[width =0.25\linewidth]{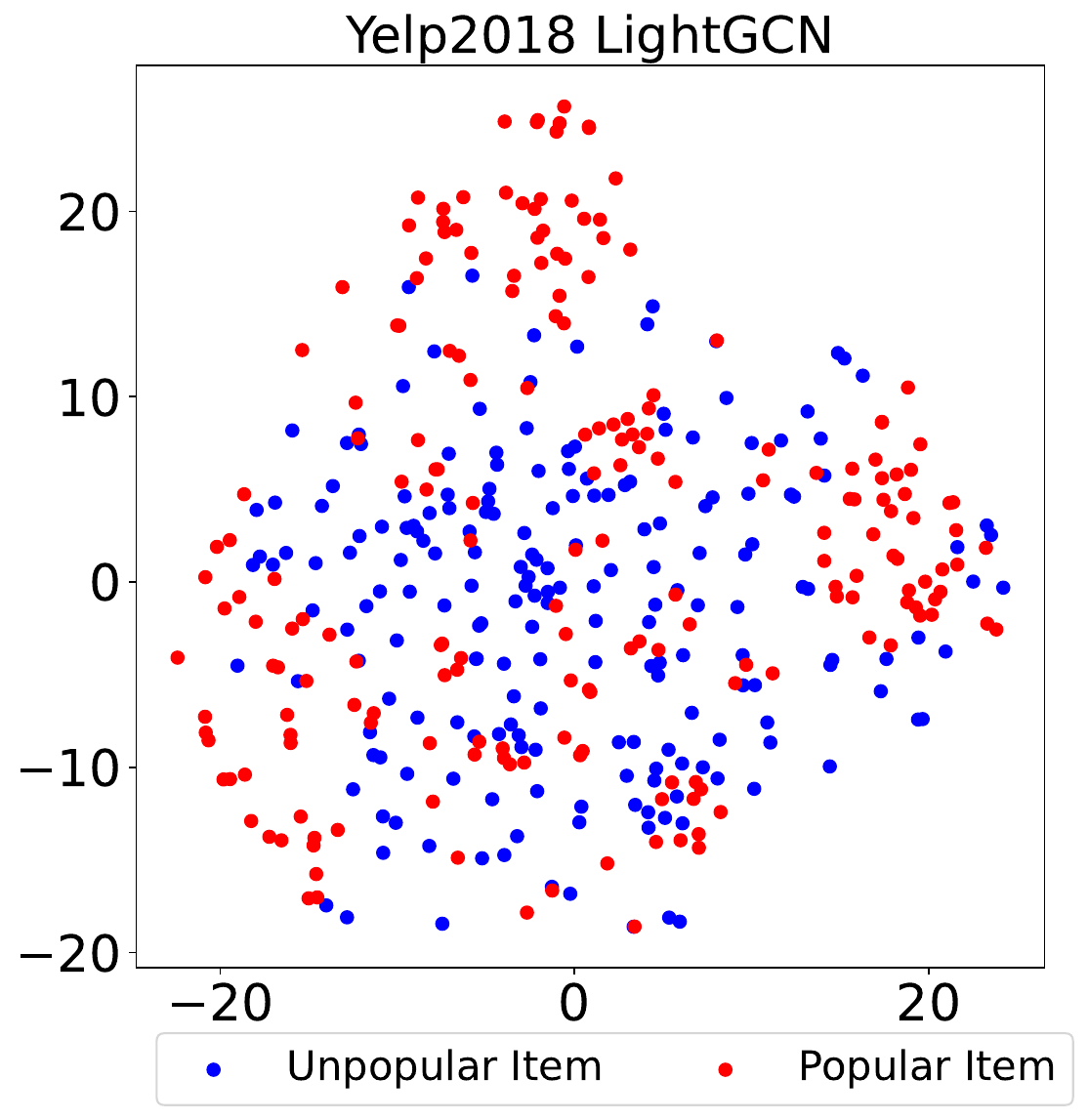}\label{fig:a_lgcn_yelp}}
    \hfill
    \subfloat{\includegraphics[width =0.25\linewidth]{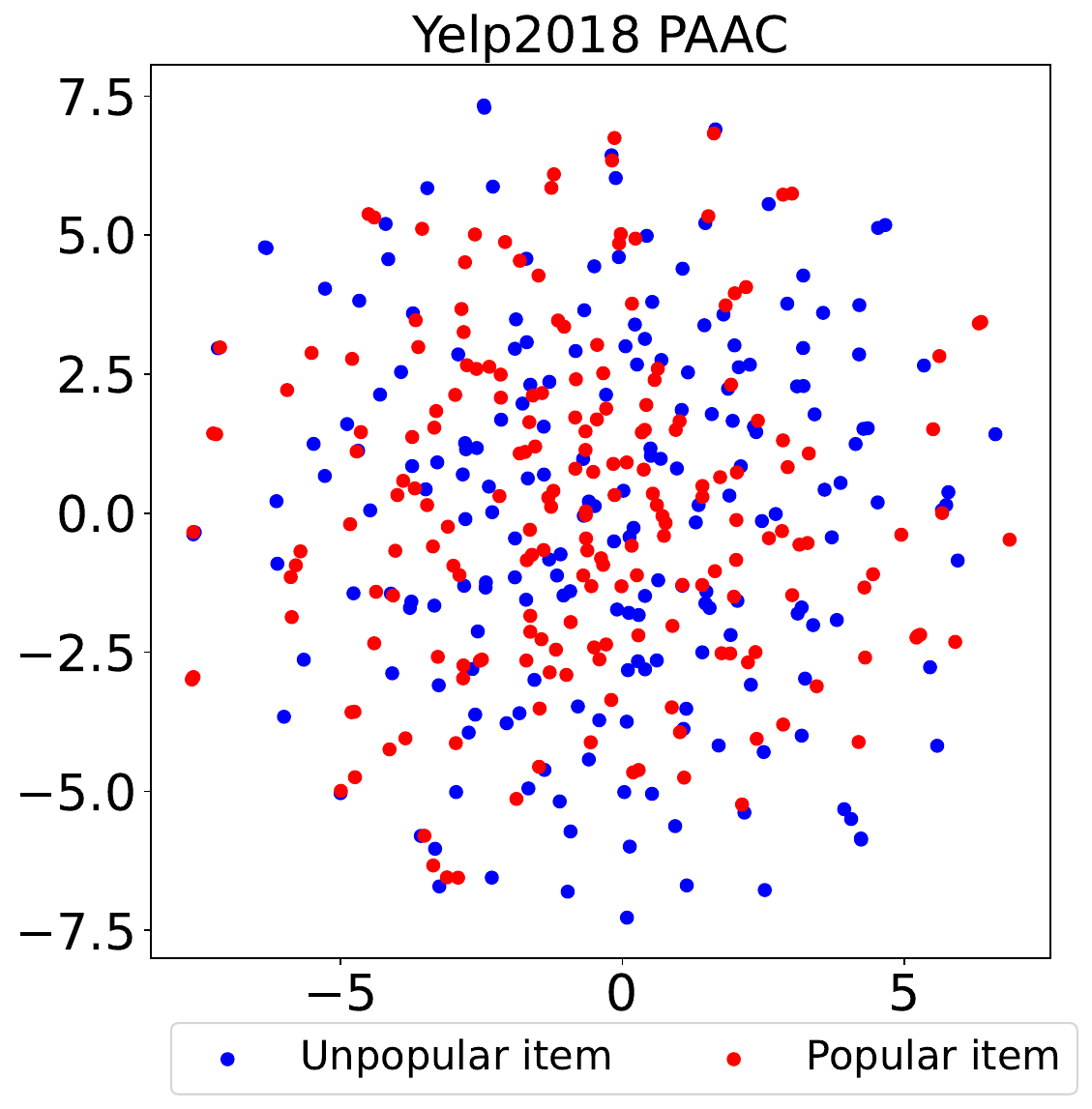}\label{fig:a_our_yelp}}
    \caption{Visualization of Item Embeddings with t-SNE~\cite{van2008visualizing}: We present a t-SNE visualization of 400 randomly selected item embeddings from our dataset.}
    \label{fig:gowalla_visual}
\end{figure*}

\textbf{Quantitative Analysis}. Following the Pareto principle~\cite{yang2022hicf}, items are categorized into 'Popular' and 'Unpopular' groups. To quantify the differences in the distribution of embeddings between these groups, we employed two key metrics: Maximum Mean Discrepancy (MMD) and Cosine Similarity (CS). MMD measures the statistical distance between two distributions and is defined as:
\[
\text{MMD}^2(X, Y) = \left\|\frac{1}{m}\sum_{i=1}^m \phi(x_i) - \frac{1}{n}\sum_{j=1}^n \phi(y_j)\right\|^2,
\]
where $X$ and $Y$ represent samples from the two distributions, with $m$ and $n$ as their respective sample sizes, and $\phi$ denotes a feature map into a reproducing kernel Hilbert space. Conversely, Cosine Similarity (CS) evaluates the angular difference between vectors, reflecting the diversity in their representations:
\[
\text{CS}(x_1, x_2) = \frac{x_1 \cdot x_2}{\|x_1\| \|x_2\|},
\]
where $x_1$ and $x_2$ are the embedding vectors. Table.~\ref{tab:quantitative} shows the results for MMD and CS, demonstrating that ~\shortname ~ significantly lowers the disparity between the distributions of popular and unpopular items—as evidenced by reduced values in both metrics—thereby effectively enhancing embedding separation.

\textbf{Qualitative Analysis}. The effectiveness of ~\shortname ~ is further validated through a comprehensive visual assessment using t-SNE~\cite{van2008visualizing}, a widely used technique for dimensionality reduction and visualization. This analysis involved visualizing the embeddings of a randomly selected subset of 400 items from the dataset. By plotting these embeddings, we are able to visually compare the performance of ~\shortname ~ against a baseline model, LightGCN.

Figure~\ref{fig:gowalla_visual} provides a clear illustration of this comparison. The figure demonstrates that PAAC achieves a significantly more uniform distribution of embeddings compared to LightGCN. In particular, both popular and unpopular items are more evenly dispersed throughout the embedding space under ~\shortname, rather than clustering in separate regions. This uniform distribution is indicative of reduced embedding separation, a crucial factor in mitigating popularity bias. The visual assessment, therefore, provides strong qualitative evidence that complements our statistical analyses, reinforcing the overall robustness and efficacy of ~\shortname ~ in promoting fairer and more balanced representation of items in the embedding space.

\end{document}